\documentstyle[preprint,prb,aps,epsf,eqsecnum]{revtex}
\begin{document}
\tightenlines

\title{Scattering of $^3$He Atoms from $^4$He Surfaces}
\author{E.~Krotscheck and R. ~Zillich}
\address{Institut f\"ur Theoretische Physik,
Johannes Kepler Universit\"at, A 4040 Linz, Austria}
\maketitle

\begin{abstract}

We develop a first principles, microscopic theory of impurity atom
scattering from inhomogeneous quantum liquids such as adsorbed
films, slabs, or clusters of $^4$He. The theory is built upon a
quantitative, microscopic description of the ground state of both the
host liquid as well as the impurity atom. Dynamic effects are
treated by allowing all ground--state correlation functions
to be time--dependent.

Our description includes both the elastic and inelastic coupling of
impurity motion to the excitations of the host liquid. As a specific
example, we study the scattering of $^3$He atoms from adsorbed
$^4$He films. We examine the dependence of ``quantum reflection''
on the substrate, and the consequences of impurity bound states,
resonances, and background excitations for scattering properties.

A thorough analysis of the theoretical approach and the physical
circumstances point towards the essential role played by inelastic
processes which determine almost exclusively the reflection
probabilities. The coupling to impurity resonances within the film
leads to a visible dependence of the reflection coefficient on the
direction of the impinging particle.

\end{abstract}
\section{Introduction}
\label{intro}

Dynamic scattering processes of helium atoms from low temperature
liquid $^4$He films and the bulk fluid in the vicinity of a free
surface continue to be a subject of considerable interest.
Experimental information is available mostly for $^4$He scattering
processes, connected with quantum reflection and quantum evaporation
\cite{WyattReflectivity2,WyattReflectivity1,WyattSpectrum,WyattRough},
as well as the surface reflectivity
\cite{Edwards75,EdwardsFatourosScattTh,Nayak83,%
SwansonEdwardsScatt4He,WyattReflectivity2}. Due to experimental
difficulties, there are only few data for $^3$He
scattering\cite{EdwardsFatourosHe3Scatt}, but there is also
interest
(experimental \cite{Godfried85,Berkhout86,Doyle91,YuDoyle93}, and theoretical
\cite{ColeSticking,ColeQuantumReflection,Bittner94} ) in the dynamics of atomic
Hydrogen atoms on $^4$He surfaces for which our theory also applies.

This paper follows up on a line of work studying the properties and
the dynamic features of quantum liquid films from a manifestly
microscopic point of view. Most relevant for the present work are
papers designing the theory for the background host
liquid\cite{filmstruc}, its excitations\cite{filmdyn,filmexc}, and the
dynamics of atomic impurities\cite{filmdirt}. In that work, we have
used the method of correlated variational wave functions which has in
many situation proven to be a computationally efficient, precise, and
robust method for the purpose of studying strongly interacting quantum
liquids. Even the simplest approximation of the theory has in the past
given quite satisfactory results on the nature of the impurity
states\cite{Impu}, their effective mass\cite{KSE} and the
impurity-impurity interaction\cite{EKS} in inhomogeneous geometries.
The reason for the {\it qualitative\/} success of the theory is that
it contains a consistent treatment of both the short- and the
long-range structure of the system. This implies that both the low-
and the high-lying excitations are treated accurately.

The present paper complements a similar study of the scattering of
$^4$He atoms from $^4$He slabs\cite{scatlett}; the problem at hand is
somewhat simpler since there is no need to fully symmetrize the wave
function of the background system {\it and\/} the impinging particle.
Another major physical difference to the scattering of $^4$He
particles is that in the latter case one might
observe\cite{HalleyDrops,ChuckWoodsLT20,SettyHalleyCampbell97} the
coupling to the Bose-Einstein condensate, whereas in the present case
one can couple both to phonon-like and to single particle
excitations. Nevertheless we will see that many similarities exits
between the two problems: The scattering process is dominated by
inelastic channels, mostly the coupling to ripplonic excitations.

Generally, the impinging particle can, in the presence of other
particles like the film of $^4$He under consideration here, scatter
into three types of channels:
\begin{enumerate}
\item
Elastic reflection: The incoming particle, characterized by the
wave vector $({\bf k}_\|,k_\perp)$, is elastically reflected with a
probability $|R|^2$. It creates virtual excitations of the background,
but transfers no energy.
\item
Inelastic scattering: with a probability $r_{\rm inel}$ the particle
loses some energy to an excitation of the film, and retains enough
energy to leave the attractive potential of the film and the
substrate.  The film excitation can be either a collective wave
(ripplon, phonon), or a single $^4$He that is elevated above the
chemical potential $\mu_4$ and leaves the film. The creation of
several excitations is in principle also included in our theoretical
description, but it is ignored in the linearized treatment of the
equations of motion.
\item
Adsorption: as in the previous case, the film is excited, but
the particle is adsorbed to the film. The corresponding {\it sticking
coefficient\/} $s$ is the probability for this process.
\end{enumerate}

\begin{figure}[h]
 \epsfxsize=12truecm
\centerline{\epsffile{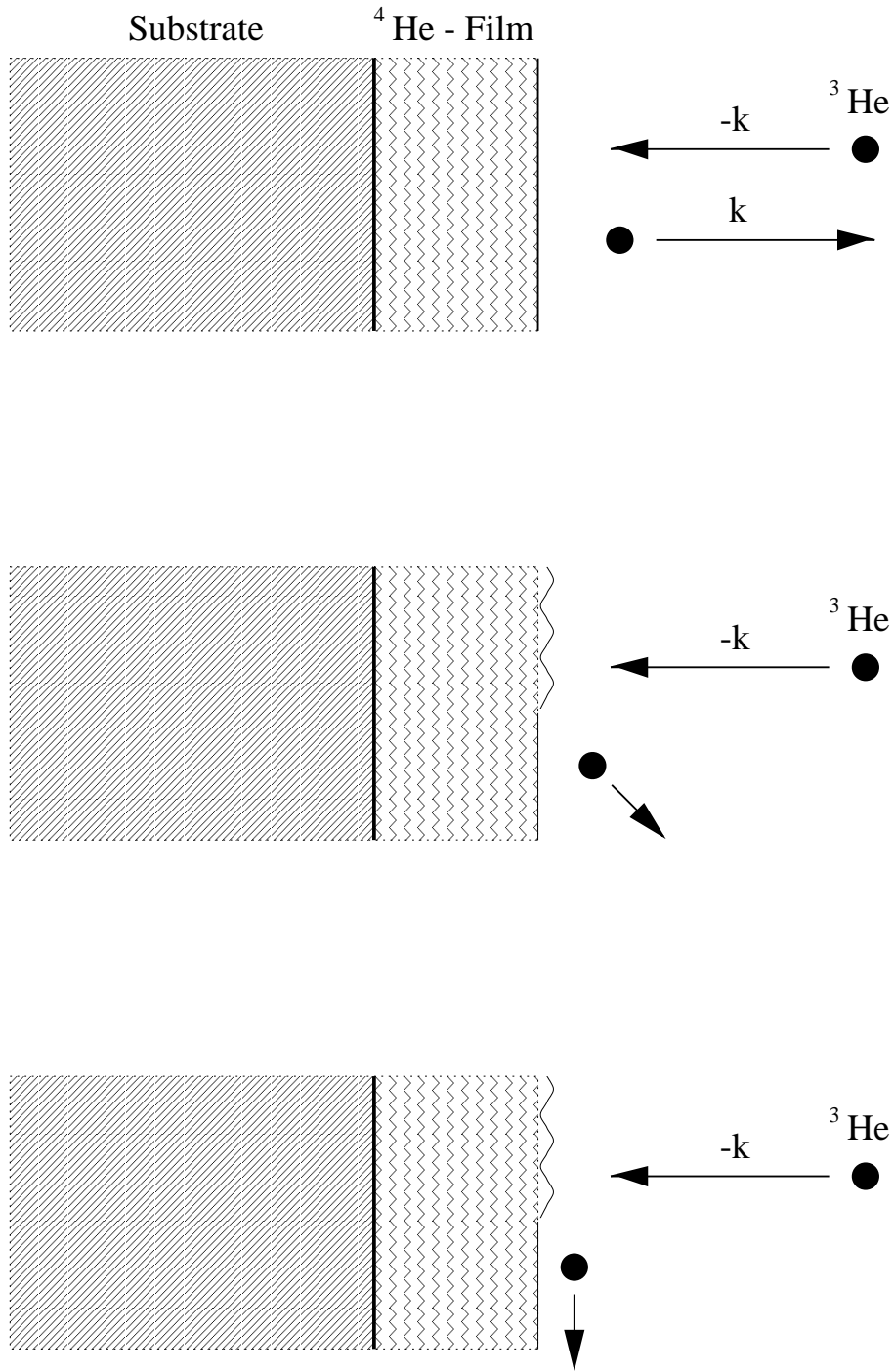}}
\vspace {1truecm}
 \caption[Erkl"arung zu Streukan"alen] {
        \label{FIGstreukanal}
        The three classes of scattering channels are illustrated.
        The incoming particle can be (a) scattered elastically
        (top figure),
        (b) inelastically, (middle figure), or (c) adsorbed to the
        film (bottom figure).
 }
\end{figure}

These three types of processes are depicted in Fig. \ref{FIGstreukanal}.
Because of the hermiticity of the {\em many--body\/} Hamiltonian
for $N$ $^4$He atoms and the $^3$He impurity, we have
\begin{equation}
|R|^2 + r_{\rm inel} + s = 1.
\end{equation}

This work focusses on the calculation of elastic scattering because
the impinging particle couples, in particular at low energies,
predominantly to the low--lying, bound excitations of the background
film and the impurity atom. We shall argue below that, basically
for phase-space reasons, inelastic processes are expected to be less
important than either elastic, or total absorption processes.

Since most of the theoretical tools of the present study have been
derived in Ref. \onlinecite{filmdirt}, we outline in
Sec. \ref{Theory} only briefly the theoretical methods and the basic
equations to be solved. The scattering problem will be be formulated
in terms of a non-local, energy dependent ``optical potential'' which
depends explicitly on the coupling of the impinging particle to
background and impurity excitations.

The results of our calculations are discussed in section
\ref{results}. To cover a variety of physical situations,
we will present results for several of the systems that were studied
extensively in our previous calculations: These will range from
strongly bound films on a model graphite substrate that is covered
with two layers of solid helium, to a very weakly bound model,
described by a rather thick, metastable film on a Cesium substrate.  We
first discuss the possible excitations of the background systems, and
then present results for the surface reflectivity as a function of
impact energy and angle for some of those systems. At very low
energies, we will encounter the effect of ``quantum reflection''
\cite{KohnSticking,BrenigSticking,BrenigSticking2,Bittner94,%
Brivio94,ColeSticking,ColeQuantumReflection};
with increasing impact energies we also can analyze the
influence of surface excitations (ripplons) and the Andreev state,
phonon/roton creation, and under certain circumstance the coupling to
an ``Andreev resonance'' of the impurity particle close to the
substrate.

\section{Microscopic Theory}
\label{Theory}
The theoretical description of $^4$He films and impurity properties
starts with a description of the ground state of the background
system. Next, a single impurity is added, and finally this impurity is allowed
to move. The technical derivation and in particular the important
verification of our theoretical tools have been presented in a series
of previous papers\cite{OldOldSurf,filmstruc,filmdirt}, we
will therefore discuss the theoretical background only briefly.

\subsection{The Background Liquid}
In the first step, one calculates the properties of the background
helium film. The only phenomenological input to the theory is the
microscopic Hamiltonian
\begin{equation}
        H_N = \sum_{1\le i\le N}\left[-{\hbar^2\over 2m_B}\nabla_i^2
        + U_{\rm sub}({\bf r}_i)\right]
        + \sum_{1\le i<j\le N} V(|{\bf r}_i-{\bf r}_j|)\ ,
\label{hamiltonian}
\end{equation}
where $V(|{\bf r}_i-{\bf r}_j|)$ is the $^4$He-$^4$He interaction, and
$U_{\rm sub}({\bf r})$ is the external ``substrate'' potential. The
many--body wave function is modeled by the {\it Jastrow-Feenberg
ansatz\/}
\begin{equation}
        \Psi_N({\bf r}_1,\ldots,{\bf r}_N)
        = \exp {1\over2}\Biggr[\sum_{1\le i\le N} u_1({\bf r}_i)
        + \sum_{1\le i<j\le N} u_2({\bf r}_i,{\bf r}_j)
        + \sum_{1\le i<j<k\le N}u_3({\bf r}_i,{\bf r}_j,{\bf r}_k)\Biggr]\ .
\label{wavefunction}
\end{equation}
An essential part of the method is the {\it optimization\/} of the
many-body correlations by solving the Euler equations
\begin{equation}
        {\delta E_N\over \delta u_n}({\bf r}_1,\ldots,{\bf r}_n) = 0
        \qquad(n = 1,\, 2,\, 3)\ ,
\label{euler}
\end{equation}
where $E_N$ is the energy expectation value of the $N$-particle
Hamiltonian (\ref{hamiltonian}) with respect to the wave function
(\ref{wavefunction}),
\begin{equation}
        E_N = {\int d^3r_1\ldots d^3r_N
        \Psi_N({\bf r}_1,\ldots,{\bf r}_N) H_N
        \Psi_N({\bf r}_1,\ldots,{\bf r}_N)\over
        \int d^3r_1\ldots d^3r_N\Psi_N^2({\bf r}_1,\ldots,{\bf r}_N)}\ . 
\label{energy}
\end{equation}
The energy is evaluated using the hyper-netted chain (HNC) hierarchy of
integral equations\cite{FeenbergBook}; ``elementary diagrams'' and
triplet correlations have been treated as described in Ref.
\onlinecite{filmstruc}.

The HNC equations also provide relationships between the {\it
correlation functions\/} $u_n({\bf r}_1,\ldots,{\bf r}_n)$ and the
corresponding $n$-body densities. One of the quantities of primary
interest is the pair distribution function $g({\bf r}_1,{\bf r}_2)$
and the associated (~real-space~) static structure function
\begin{equation}
        S({\bf r}_1,{\bf r}_2) = \delta({\bf r}_1-{\bf r}_2)
        +\sqrt{\rho_1({\bf r}_1)\rho_1({\bf r}_2)}
        [g({\bf r}_1,{\bf r}_2)-1]\,.
\label{static}
\end{equation}
The static structure function and the effective
one-body Hamiltonian
\begin{equation}
H_1({\bf r})    = -{\hbar^2\over 2m_B}
        {1\over\sqrt{\rho_1({\bf r})}}\nabla\rho_1({\bf r})
        \nabla{1\over\sqrt{\rho_1({\bf r})}}
\label{hone}
\end{equation}
define the {\it Feynman excitation spectrum\/} through the
generalized eigenvalue problem
\begin{equation}
        H_1({\bf r}_1)\psi^{(\ell)}({\bf r}_1)
        = \hbar\omega_\ell\int d^3r_2 
        S({\bf r}_1,{\bf r}_2)\psi^{(\ell)}({\bf r}_2)\ ,
\label{RPA}
\end{equation}
which is readily identified with the inhomogeneous
generalization\cite{ChangCohen} of the well-known Feynman dispersion
relation\cite{Feynman} $\hbar\omega(k) = \hbar^2 k^2/2m_B S(k)$. The states
$\psi^{(\ell)}({\bf r})$, their associated energies
$\hbar\omega_\ell$, and the adjoint states
\begin{equation}
\phi^{(\ell)}({\bf r}) = {1\over\hbar\omega_\ell}
H_1({\bf r})\psi^{(\ell)}({\bf r})
\label{phidef}
\end{equation}
are useful quantities for the impurity problem and for the
representation of the dynamic structure function of the background
film.

\subsection{The Static Impurity Atom}

The Hamiltonian of the $N+1$ particle system consisting of $N$
$^4$He atoms and one impurity is

\begin{equation}
        H_{N+1}^I =-{\hbar^2\over 2m_I}\nabla_0^2 
        + U^I_{\rm sub}({\bf r}_0)
        + \sum_{i=1}^N V^I(|{\bf r}_0-{\bf r}_i|) 
        + H_N
\label{hamimptonian}
\end{equation}
We adopt the convention that coordinate ${\bf r}_0$ refers to the
impurity particle and coordinates ${\bf r}_i$, with $i=1\ldots N$ to
the background particles.  Note that the substrate potentials $U_{\rm
sub}({\bf r}_i)$ and $U^I_{\rm sub}({\bf r}_0)$, as well as the
interactions $V^I(|{\bf r}_0-{\bf r}_j|)$ and $V(|{\bf r}_i-{\bf r}_j|)$,
can be different functions for different particle species.

The generalization of the wave function (\ref{wavefunction}) for an
inhomogeneous $N$-particle Bose system with a single impurity atom is
\begin{eqnarray}
&&      \Psi_{N+1}^I({\bf r}_0,{\bf r}_1,\ldots,{\bf r}_N)
\label{impwave}\\
&&      = \exp{1\over2}\biggl[u^I_1({\bf r}_0)
        +\sum_{1\le i\le N} u^I_2({\bf r}_0,{\bf r}_i)
        +\sum_{1\le i<j\le N} u^I_3({\bf r}_0,{\bf r}_i,{\bf r}_j)
\biggr]\Psi_N({\bf r}_1,\ldots,{\bf r}_N)\ .
\nonumber
\end{eqnarray}

The energy necessary for (or gained by) adding one
impurity atom into a system of $N$ background atoms is the impurity
chemical potential
\begin{equation}
\mu^I\equiv E_{N+1}^I-E_N.
\label{chemI}
\end{equation}
Here, $E_{N+1}^I$ is to be understood as the energy expectation value
of the Hamiltonian (\ref{hamimptonian}) with respect to the wave
function (\ref{impwave}). The further steps parallel those of
the derivation of the background structure.

The impurity density is calculated by minimizing the chemical potential
(\ref{chemI}) with respect to $\sqrt{\rho^I_1({\bf r}_0)}$. This leads
to an effective Hartree equation
\begin{equation}
        -{\hbar^2\over 2m_I}\nabla^2_0\eta^{(r)}({\bf r}_0)
        +\bigl[U^{\rm I}_{\rm sub}({\bf r}_0)+V_H({\bf r}_0)\bigr]
        \eta^{(r)}({\bf r}_0) = t_r \eta^{(r)}({\bf r}_0) 
\label{HARimp}
\end{equation}
where $V_H({\bf r}_0)$ is an effective, self-consistent one-body
potential for the single impurity. The lowest eigenvalue of
Eq. (\ref{HARimp}) is the impurity chemical potential $\mu^I =
t_0$, and the corresponding eigenfunction the density of the
impurity ground state, $\sqrt{\rho^I_1({\bf r})} = \eta^{(0)}({\bf
r})$.

In the systems studied below, translational invariance in the $x-y$
plane is assumed, and the states are characterized by two quantum
numbers, $m$ and ${\bf k}_m$, associated with the motion perpendicular
($m$) and parallel to the symmetry plane (${\bf k}_m$). When
unambiguous, as in the states $\eta^{(r)}({\bf r}_0)$ and
$\phi^{(m)}({\bf r}_1)$, we shall use the single label (e.g. $m$) to
collectively represent both quantum numbers. In particular, the states
$\eta^{(r)}({\bf r}_0)$ depend only trivially on the parallel
coordinate,
\begin{equation}
        \eta^{(r)}({\bf r}_0) = \eta^{(r)}(z_0) 
        e^{i{\bf k}_r\cdot{\bf r}_\|}\ .
\end{equation}
The unit volume is chosen as the size of the normalization volume.
The corresponding energies are
\begin{equation}
        t_r = \epsilon_r + {\hbar^2 k_\|^2\over 2m_I}\ ,
\label{bareenergy}
\end{equation}
where $\epsilon_r$ are the eigenvalues of Eq. (\ref{HARimp}) for $k_\|
= 0$.

\subsection{Impurity dynamics}
\label{timedep}

It is tempting to identify the higher-lying eigenstates of the
``Hartree-equation'' (\ref{HARimp}) with the excited states of the
impurity. This is legitimate only in a {\it static approximation\/}
for the impurity features. However, such a simplification misses two
important features:

\begin{itemize}

\item{} If the momentum is a good quantum number, low-lying excited
states can be discussed in terms of an {\it effective
mass.\/} In our geometry, a ``hydrodynamic effective mass'' is
associated with the motion of an impurity particle {\it parallel\/} to
the surface; it is caused by the coupling of the impurity
motion to the excitations of the background liquid. The
local Hartree--equation (\ref{HARimp}) misses this effect.

\item{} The effective Hartree-potential $V_H(z)$ is real, {\it i.e.\/}
all ``excitations'' defined by the local equation (\ref{HARimp}) have
an infinite lifetime. A more realistic theory should
describe resonances and allow for their decay by the coupling to the
low-lying background excitations of the host film.

\end{itemize}

Hence, a static equation of the type (\ref{HARimp}) is appropriate for
the impurity ground state only.  The natural generalization of the
variational approach to a dynamic situation is to allow for
time--dependent correlation functions $u_n({\bf
r}_0,\ldots,{\bf r}_n;t)$. We write the time dependent variational
wave function in the form
\begin{equation}
        \phi (t) =
        {1\over \sqrt{\left\langle\psi^I\mid\psi^I\right\rangle}}
        e^{-iE_{N+1}^I t / \hbar }
        \psi^I ({\bf r}_0,{\bf r}_1,...{\bf r}_N;t)\ .
\label{JFimpuDyn}
\end{equation}
Consistent with the general strategy of variational methods, we
include the time dependence in the one-particle {\it and\/}
two-particle impurity-background correlations, {\it i.e.\/} we write
\begin{equation}
        \psi^I({\bf r}_0, {\bf r}_1,... {\bf r}_N;t)
        = \exp {1 \over 2}
        \Bigl[\delta u_1({\bf r}_0;t) +  \sum_{1\le i\le N}
                \delta u_2({\bf r}_0,{\bf r}_i;t) \Bigr]
        \Psi^I_{N+1}({\bf r}_0,{\bf r}_1,...,{\bf r}_N)\ .
\end{equation}
The time independent part remains the same as defined in
Eq.~(\ref{impwave}). The time--dependent correlations are determined
by searching for a stationary state of the action integral
\begin{eqnarray}
        {\cal S} &&= \int^t_{t_0}{\cal L}(t) dt
\nonumber\\
{\cal L}(t) &&= \langle \phi (t) \vert H_{N+1}^I-i\hbar {\partial \over
                \partial t} \vert \phi (t) \rangle\ ,
\label{S}
\end{eqnarray}
where $H_{N+1}^I$ is the Hamiltonian (\ref{hamimptonian}) of the
impurity-background system.

The derivation of a set of useful equations of motion for the impurity
have been given in Ref. \onlinecite{filmdirt}.  The final result is
readily (and expectedly) identified with a Green's function
expression, where the three-body vertex function describes an impurity
atom scattering off a phonon, and is given in terms of quantities
calculated in the ground-state theory. The motion of the impurity
particle is determined by an effective Schr\"odinger equation of the
form
\begin{equation}
\left[-\frac{\hbar^2}{2m_I}\nabla^2
+U_{\rm sub}^I+V_H({\bf r})\right]\psi_I({\bf r},\omega)
+ \int d^3r'\,\Sigma({\bf r},{\bf r'},\omega)\psi_I({\bf r'},\omega)
=\hbar\omega\psi_I({\bf r},\omega)\,,
\label{onedyn}
\end{equation}
where  $V_H({\bf r})$ is the effective one--body potential
of Eq. (\ref{HARimp}), and $\Sigma({\bf r},{\bf r'},\omega)$
is the impurity self--energy. Within the chosen level of the
theory, $\Sigma({\bf r},{\bf r'},\omega)$ is describes
three--body processes,
\begin{equation}
        \Sigma({\bf r},{\bf r'},\omega)
        = \sum_{rm}{W_{mr}({\bf r})W_{mr}({\bf r'})\over
        \hbar\omega - \hbar\omega_m - t_r}\,,
\label{selfimp}
\end{equation}
where $W_{mr}({\bf r})$ is the three--body vertex function that
describes the coupling between an incoming $^3$He particle to an
outgoing $^3$He in the state $r$ as well as an outgoing phonon
in state $m$. The detailed form of these matrix elements
follows from the microscopic theory that has been described in length
in Ref. \onlinecite{filmdirt}, it is not illuminating for the further
considerations.

The structure of Eqs. (\ref{onedyn}) and (\ref{selfimp}) is of the
expected form of an energy-dependent Hartree-equation with a
self-energy correction involving the energy loss or gain of the
impurity particle by coupling to the excitations of the background
system. It is the simplest form that contains the desired
physical effects.

The energy denominator in Eq.  (\ref{selfimp}) contains the Feynman
excitation energies defined in Eq. (\ref{RPA}) and the Hartree
impurity energies of Eq. (\ref{HARimp}). These energies are too high,
and we expect therefore that three--body effects are somewhat
underestimated.  A lowering of the spectra in the energy denominator
by an impurity effective mass or by a more quantitative phonon/roton
spectrum should have the effect of enhancing the importance of
multi-particle scattering processes. Hence, it is expected that the
binding energy of the surface resonance is still somewhat too high
compared with experiments. On the other hand, it is not expected that
a more quantitative spectrum in the self--energy should change the
effective mass of the Andreev state considerably because the
hydrodynamic backflow causing this effective mass is mostly caused by
the coupling to ripplons, which are well described within the Feynman
approximation.

\section{The physical models}
\label{Model}

We consider liquid $^4$He adsorbed to a plane attractive substrate
which is translationally invariant in the $x-y$ plane, {\it i.e.}
$U_{\rm sub}({\bf r}) = U_{\rm sub}(z)$. The systems under
consideration are characterized by the substrate potential $U_{\rm
sub}(z)$ and the surface coverage
\begin{equation}
        n = \int_0^\infty dz \rho_1(z)\ ,
\label{coverage}
\end{equation}
where $\rho_1({\bf r}) = \rho_1(z)$ is the density profile of the
$^4$He host system. This density profile is, along with the
energetics, structure functions, and excitations of the film, obtained
through the optimization of the ground-state (\ref{wavefunction}) as
outlined above; the procedure has been described in detail in
Ref. \onlinecite{filmstruc}.

\subsection{Ground state}

\begin{figure}[h]
 \epsfxsize=15truecm
\centerline{\epsffile{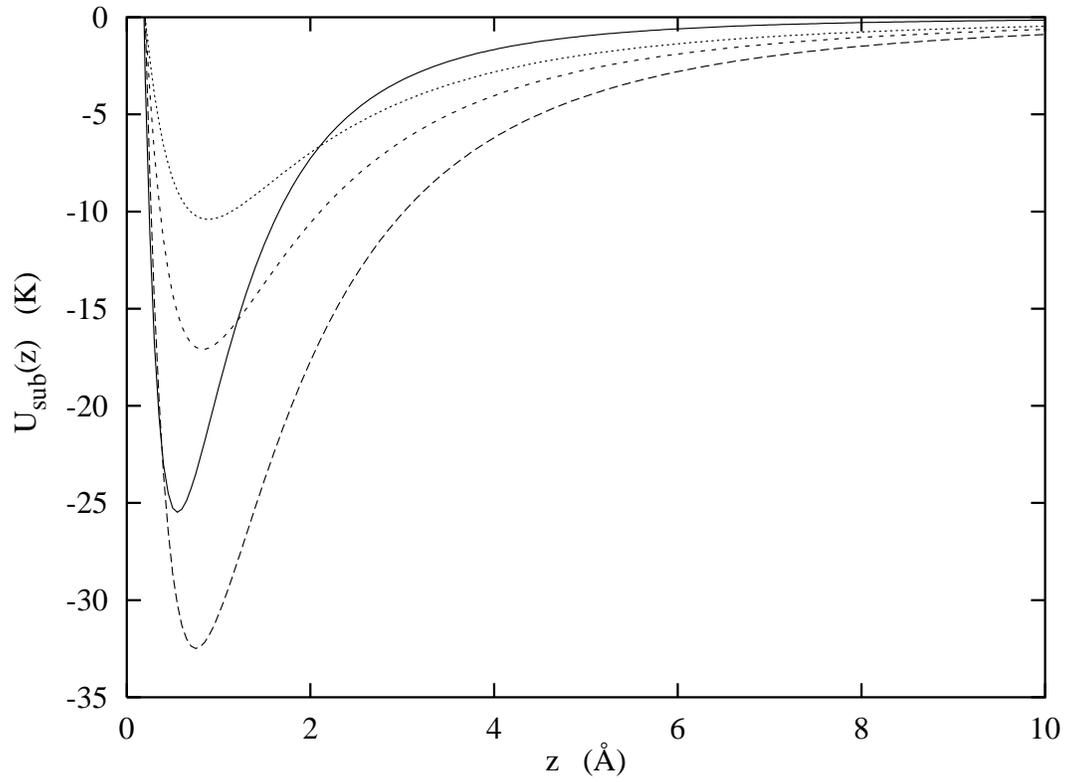}}
\vspace{2truecm}
\caption{
The figure shows the three substrate potentials for the films
under consideration here: Graphite {\it plus\/} two solid helium
layers (solid line), Mg (long dashed line), Na (short
dashed line) and Cs (dotted line).\label{fig:uexts}}
\end{figure}

\begin{figure}[h]
 \epsfxsize=15truecm
\centerline{\epsffile{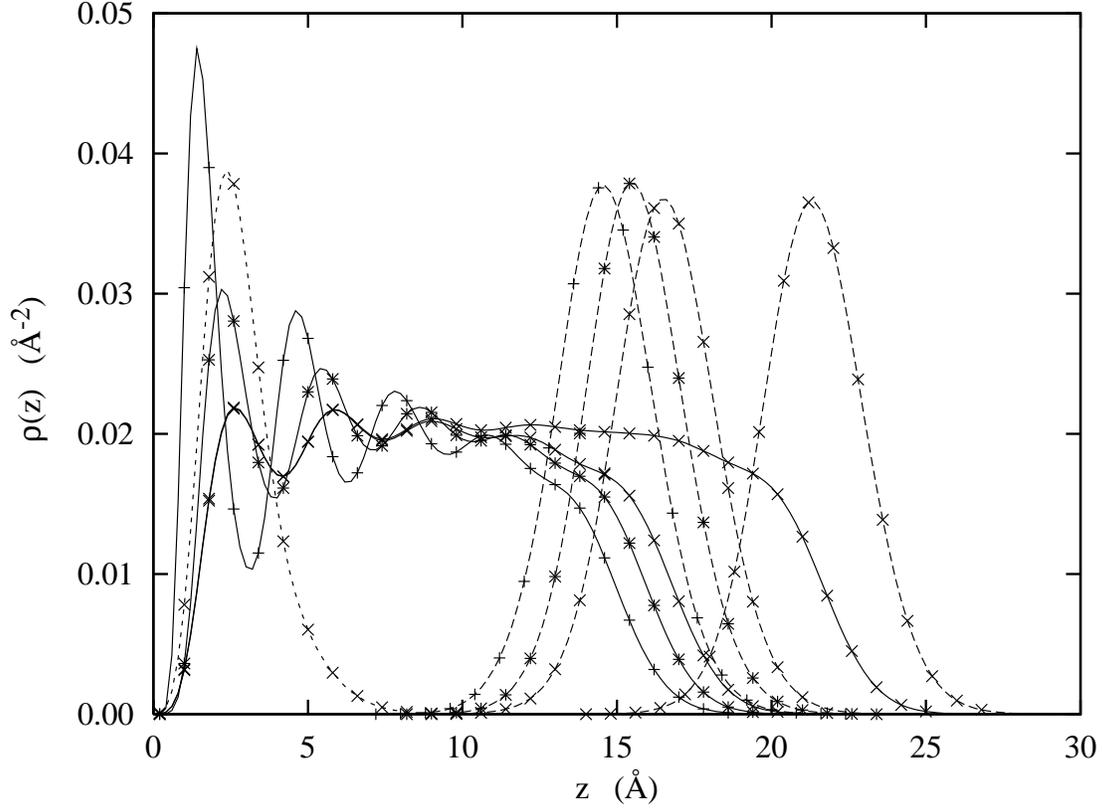}}
\vspace{2truecm}
 \caption{The figure shows the four density profiles of the background
liquid (solid lines) and the impurity location (long dashed lines)
for which most of the present calculations were done. Graphite
substrate results are marked with $+$-symbols, Na results with
stars, and Cs results with crosses. Also shown is the interfacial
Andreev state on a Cs substrate (short-dashed line marked with crosses).
Coverages are $n = 0.30~{\rm\AA}^{-2}$ for Cs, Na, and graphite,
and $n = 0.40~{\rm\AA}^{-2}$ for Cs. Profiles on Mg have been left
out for clarity.
\label{fig:scprof}}
\end{figure}

We have in this work studied the scattering properties of $^4$He atoms
for a number of selected substrate potentials and surface coverages;
we have selected four cases for the purpose of a detailed discussion.
The substrate potentials and the corresponding density profiles are
shown in Figs. \ref{fig:uexts} and \ref{fig:scprof}. The surface
coverages are $n = 0.3\,{\rm\AA}^{-2}$ for each substrate potential;
additionally we have considered the case $n = 0.4\,{\rm\AA}^{-2}$ for a
Cs substrates as well as Mg for a case that is somewhat more
attractive than the screened graphite, but also has a long range.

Alkali metal substrate potentials are simple $3-9$ potentials
characterized by their {\it range\/} $C_3$ and their {\it well
depth\/} $D$. They have the form
\begin{equation}
U_{\rm sub}(z) = \left[{4C_3^3\over 27D^2}\right]{1\over z^9}
 - {C_3\over z^3}.
\label{Usubst}
\end{equation}
The range parameters $C_3$ of these potentials have been calculated by
Zaremba and Kohn\cite{ZarembaKohn77}, the short--range $z^{-9}$ term is
phenomenological and fitted to reproduce the binding energies of a
single atom on these substrates. Slightly more complicated is our
model of a graphite substrate covered with two solid layers of
$^4$He. Most important for low--energy scattering properties is the
coefficient $C_3$ of the long--range attraction, the values of $C_3$
for our substrates of graphite, Cs, Na, and Mg are $180$, $ 670$,
$1070$ and $1750$~K~${\rm\AA}^3$, respectively\cite{ZarembaKohn77,Cole}.
The graphite-- potential is relatively short-range but deep and
produces a very visible layering structure of the background film;
thus one obtains a rather ``stiff'' system\cite{filmstruc}.

Fig. \ref{fig:uexts} provides a comparison of these four different
potentials. It is seen that the alkali metal potentials are longer
ranged, the magnesium substrate has the deepest potential well.  At
the opposite end of the potential strength is the Cs substrates. This
substrate has received much attention in recent years because of the
experimental finding that it is
non-wetting\cite{Nacher,Ketola,Taborek}.  Note that the Cs-adsorbed
films are metastable; they were examined with two purposes in
mind. One is to generate a situation that is reasonably close the
infinite half--space limit. Therefore, we have studied this case also
for the larger surface coverage $n = 0.4\,{\rm\AA}^{-2}$. The second
reason is that the nature of the low--lying excitations\cite{KLW96} as
well as that of the impurity states\cite{Taborek95} is somewhat
different than those for the graphite model as will be seen below.

The third case, a Na substrate, is an intermediate case which is of
some interest for the nature of the $^3$He bound states, whereas the
Mg substrate is both deeper and longer ranged than the screened
graphite.

\subsection{Background Excitations}

Our earlier work\cite{filmdyn,filmexc,alkalis} has discussed
extensively the excitations of quantum liquid films adsorbed to
various substrates. These studies have been concerned with the
interpretation of neutron scattering
experiments\cite{LauterExeter,LauterPRL,LauterJLTP}, they have
therefore focussed on excitations propagating {\it parallel\/} to the
film. Typically, four types of modes were found:
\begin{enumerate}

\item Surface excitations: At long wavelengths and on strong
substrates, these are substrate potential driven modes
with a linear dispersion relation
\begin{equation}
\omega_3(k) = c_3 k\,,
\label{thirdsound}
\end{equation}
where $c_3$ is the speed of {\it third sound\/}. At shorter
wavelengths and in the case of an infinite half--space, the
surface--mode is driven by the surface tension and has a dispersion
relation
\begin{equation}
\omega_r^2(k) = {\sigma\over m \rho_\infty}k^3.
\label{ripplon}
\end{equation}
where $\sigma$ is the surface tension, and $\rho_\infty$ the density
of the bulk liquid. In practice, the dispersion relation is linear
only in a rather small momentum regime, and the ripplon dispersion
relation (\ref{ripplon}) is a quite good approximation\cite{alkalis}
for the surface---mode dispersion relation up to wavelengths of about
$0.5~{\rm\AA}^{-1}$. The theoretically predicted surface energy
obtained from Eq. (\ref{ripplon}) by a $k^{3/2}$ fit to the dispersion
relation is $\sigma_{\rm th} \approx 0.279~K{\rm\AA}^{-2}$ which
compares favourably with the most recent experimental
value\cite{DevilleLT21,Deville97}
of $\sigma_{\rm ex} \approx 0.279~K{\rm\AA}^{-2}$.

\item Bulk Rotons:
Films with a thickness of two or more liquid layers show already
a quite clear phonon/roton spectrum. The spectrum starts at finite
energy in the long---wavelength limit and contributes, in this
momentum regime, very little to the strength. It takes over most
of the strength in the regime of the roton minimum.

\item Layer Rotons:
Films with a strongly layered structure also show excitations
(identified as sound--like through their longitudinal current pattern)
that propagate essentially within one atomic layer. These excitations
have a two--dimensional roton with an energy {\it below\/} the bulk
roton, and have been identified with a ``shoulder'' in the neutron
scattering spectrum below the ordinary roton minimum.

\item Interfacial Ripplons: On very weak substrates, like cesium, one
can also have an ``interface ripplon\cite{KLW96,alkalis}''. Its
appearance can be understood easily from the following consideration:
Consider first a film with {\it two\/} free surfaces.  Obviously, this
film would exhibit two ripplon modes, one at each
surface\cite{Surface2}.  Now, a weak substrate is moved against one of
the two surfaces. The character of the ``ripplon'' at this surface
will not change abruptly; rather the circular motion of the particles
will be somewhat inhibited, and the energy of the mode will rise.
This is precisely what is seen in the energetics and the current
pattern of this second mode on Cs. Stronger substrate potentials
suppress this interface mode; to distinguish between an
``interfacial ripplon'' and a ``layer phonon'' one must look at the
current pattern of the excitation\cite{alkalis}.
\end{enumerate}

The above list of excitations is restricted to modes that can be
characterized legitimately by a wave vector ${\bf k}_\|$ parallel to
the surface. To calculate the response to particles impinging normally
on the surface, one must also look at the types of excitations {\it
perpendicular\/} to the surface. These cannot be rigorously classified
by a wave number, but one should basically expect standing waves or
resonances at discrete frequencies, approaching the excitations of a
bulk system as the film becomes thicker. No ripplonic excitations or
layer--modes should be visible in this case.

The character of excitations is intelligently discussed by examining
the {\it dynamic structure function\/} $S({\bf k},\omega)$. A general
procedure has been developed in Refs. \onlinecite{filmdyn,filmexc}
to use time--dependent correlations for a quantitative
calculation of the dynamic structure function. The simplest
version of the theory is analogous to the Feynman approximation
\cite{Feynman,ChangCohen}; the dynamic structure function
in that approximation can be calculated directly from the
solutions of Eq. (\ref{RPA})
\begin{equation}
        S({\bf k};\omega) = \left| \int d^3 r e^{i {\bf r}\cdot{\bf k}}
        \sqrt{\rho_1({\bf r})}\phi_\omega({\bf r})\right|^2
\end{equation}
where the $\phi_\omega({\bf r})$ are adjoint states (\ref{phidef}) of
the solutions of Eq. (\ref{RPA}) for energy $\hbar\omega$. The Feynman
approximation has its well known deficiencies, and methods for its
improvements have been derived which provide quantitative agreement
with experiments.

\begin{figure}[h]
\epsfxsize=15truecm
\centerline{\epsffile{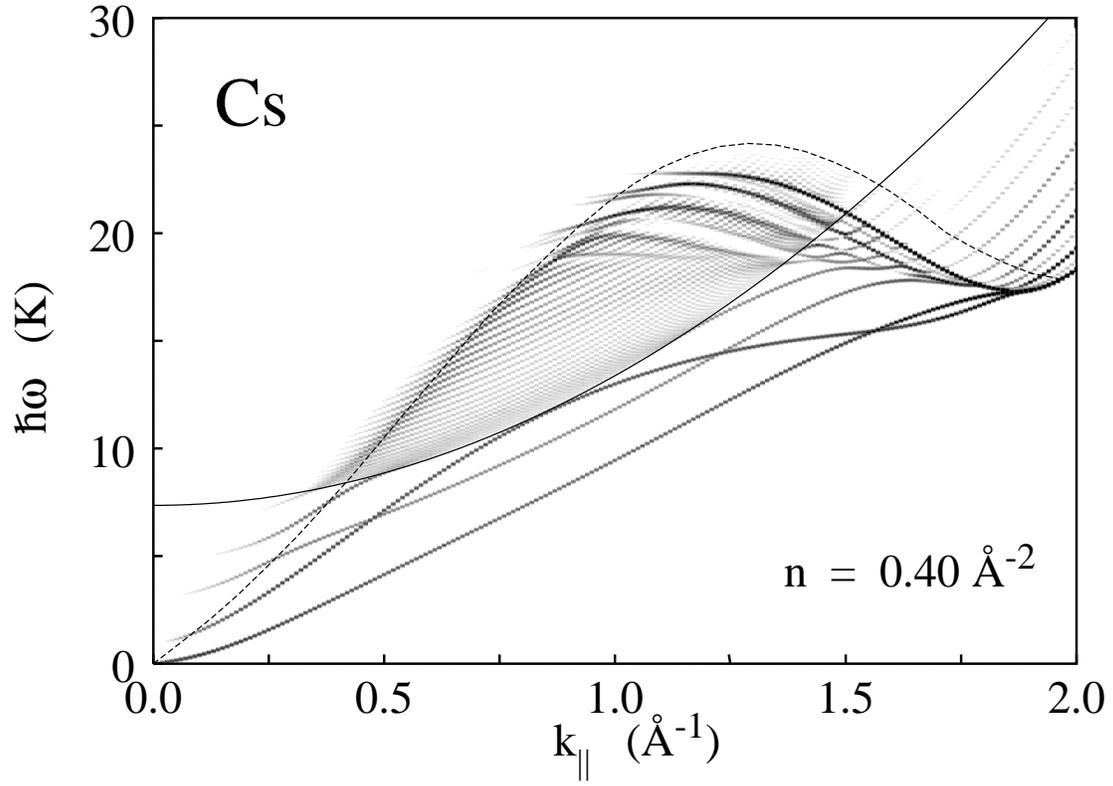}}
\vspace{2truecm}
\caption[Feynman excitations spectrum] {
\label{fig:skwcs40p}
Dynamic structure function $S(k,\omega)$ in Feynman
approximation for a film with coverage of $n=0.400{\rm \AA}^{-2}$
on a Cs substrate and {\it parallel\/} momentum transfer.
The solid curve shows the continuum boundary $-\mu + \hbar^2
q_\|^2/2m_4$ and the dashed line the bulk Feynman spectrum.}
\end{figure}

\begin{figure}[h]
\epsfxsize=15truecm
\centerline{\epsffile{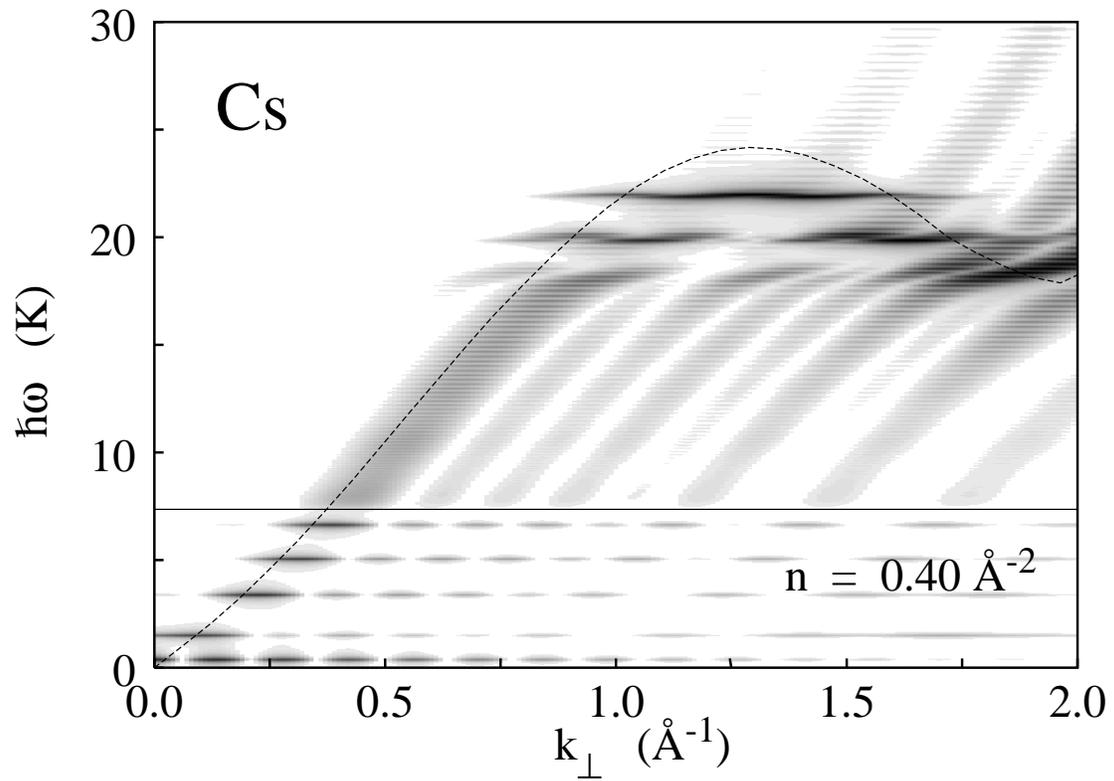}}
\vspace{2truecm}
\caption[Feynman excitations spectrum] {
\label{fig:skwcs40o}
Same as Fig. \protect\ref{fig:skwcs40p} for momentum transfer {\it
perpendicular\/} to the film. The horizontal solid line shows the
continuum boundary $-\mu$ and the dashed line the bulk Feynman
spectrum.  }
\end{figure}

Previous work has concentrated on the theoretical interpretation of
neutron--scattering experiments, it was therefore concerned with
momenta parallel to the liquid surface. In the present situation we
must allow for both parallel and perpendicular momentum transfer. We
show in Figs. \ref{fig:skwcs40p} and \ref{fig:skwcs40o} the dynamic
structure function for parallel and perpendicular momentum
transfer. Fig. \ref{fig:skwcs40p} shows the picture familiar from
previous work\cite{filmdyn,filmexc,alkalis}: a low-lying excitation
which can be identified with a ripplon by its dispersion relation and
its particle motion, and a high density of states on the roton regime;
note that the second lowest dispersion branch corresponds to the
interfacial ripplon mentioned above. Note also that the modes below
the continuum energy $-\mu_4 + \hbar^2 q_\|^2/2m_4$ are discrete; they
have been broadened by a Lorentzian of the same strength to make them
visible.

The situation is quite different for perpendicular scattering.  Again,
the discrete excitations below the the evaporation energy have been
broadened. We see a dominant ridge basically along the dispersion
relation of a Feynman phonon, and a high density of states in the
regime of the roton. The ridge shows a number of ``echoes'' at shorter
wavelengths; this is due to the finite--size of the film. But there
are --- expectedly --- no excitations corresponding to the
(interfacial) ripplons.

\subsection{Impurity Excitations}

Calculations of low-lying, bound states including the dynamic
self--energy have been discussed extensively in
Ref. \onlinecite{filmdirt}, we list here the most important ones
demonstrating both the theoretical consistency as well as the
quantitative reliability and highlight their relevance for scattering
processes:
\begin{itemize}
\item  When applied to the bulk liquid, the ground state theory
produces the correct chemical potentials of $^3$He and hydrogenic
impurities\cite{SKJLTP}.
\item  In an inhomogeneous geometry, the {\it static\/} theory
reproduces the binding energy of the Andreev state\cite{Hallock94}.
The theory also predicts, even in its most primitive
version\cite{OldOldSurf}, the existence of a surface resonance.
\item  The {\it dynamic\/} theory predicts a hydrodynamic
effective mass of the Andreev state of $m^*_H/m_I \approx 1.35$, to be
compared to the value on 1.38 given by Higley {\it at al.\/}
\cite{Hallock89} somewhat larger than the value of $m^*_H/m_I \approx
1.26$, reported by Valles {\it et al.\/} \cite{Hallock88} at the lower
end of the value $m^*_H/m_I = 1.45\pm0.1$ given by Edwards and Saam
\cite{EdSaam}.  In other words, our theoretical prediction is within
the spread of experimental values.
\item  The energy of the first excited surface state is lowered from
about -2.2~K to -2.8~K, improving the agreement with the experimental
value\cite{Hallock94} of approximately -3.2~K notably.
\end{itemize}

Similar to the obvious existence of interfacial ripplons, one also
expects, on weak substrates, the appearance of an interfacial Andreev
state. The binding energy of this state was found in
Ref. \onlinecite{filmdirt} to be approximately -4.3~K. which is
somewhat higher than the experimental value\cite{Taborek95} of
-4.8~K.  We attribute the difference to uncertainties in the substrate
potential and the certainly oversimplified assumption of a perfectly
flat surface. This state --- being confined to a smaller area than
the surface state --- has {\it always\/} an energy that is higher than
the Andreev state. Although it can in principle decay into a surface
bound state, it has negligible overlap and hence its lifetime is
practically infinite.  With increasing potential strength, the energy
of the substrate bound state increases; the state disappears
completely on substrates somewhat more attractive than Na. Then, the
``interfacial Andreev state'' turns into a resonance to which a
scattering particle can couple. Similar ``resonances'' can be found
on Mg substrates even in the {\it second\/} layer; we shall return
to this point further below.

The two surface--bound states (and, if applicable, the interfacial
Andreev state) can be described in the energy regime we are interested
in reasonably well by an $t_i(k) = t_i(0) + \hbar^2 k^2/2m_3^*$. Above
the solvation energy of a $^3$He atom, a sequence of impurity states
can exist that are spread out throughout the film; the detailed
energetics of these states depends on the thickness of the film and
the corrugation of the background liquid.

\section{Scattering states}
\label{results}

The background and impurity excitations discussed in the previous
section specify the possible energy loss channels for a scattering
particle; we can now turn to the analysis of our results.

The previous work has concentrated on the properties of {\it bound\/}
impurity atoms, their effective masses, and the lifetime of
resonances. Scattering processes are treated within the same theory,
imposing asymptotic plane--wave boundary conditions on the solution of
the effective Schr\"odinger equation (\ref{onedyn}):
\begin{equation}
\psi_{\rm I}(z,{\bf r}_\|)
\rightarrow\  e^{i{\bf k}_\|\cdot{\bf r\/}}
\left[e^{-i k_\perp z} + Re^{i k_\perp z}\right]
{\rm as}\quad z\rightarrow\infty.
\label{asymptotia}
\end{equation}

One of the key quantities of the theory is the elastic reflection coefficient
$R$ because it is directly influenced by the coupling of the motion of
the impinging particle to the excitations of the quantum liquid.  The
absolute value of the reflection coefficient can differ from unity
only if the self--energy $\Sigma({\bf r},{\bf r}',\omega)$ is
non-hermitian. This happens when the energy denominator in the
self--energy (\ref{selfimp}) has zeroes; note that the quantum numbers
$m$ and $r$ include both the motion of the particles parallel to the
surface as well as the discrete or continuous degrees of freedom in
the $z$-direction.

Superficially, we appear to be describing a single--particle quantum
mechanical scattering problem. In fact, a number of notions can be
carried over from single particle models can be carried over, and
simple phenomenological descriptions can be constructed at the level
of a one--body theory. But the actual situation is far richer: Since
the scattering film is composed of helium atoms, this is a
generically {\it non-local\/} problem when viewed at the one-body
level. Moreover, the film is {\it dynamic:} the incoming particle may
produce excited states of the background. This may result in the
capture of the particle and/or the emission of particles in states
other than the elastic channel.

\subsection{Quantum reflection}

Generally, the amplitude of the wave function of an impinging particle
of low energy is suppressed inside an attractive potential by the
mismatch of the wave lengths inside and outside the potential if its
range is small compared to the wavelength of the particle. As a
consequence, the particles are almost totally reflected even if there
is dissipation inside the potential (caused by the imaginary part of
the self energy operator (\ref{selfimp}) in our case)
\begin{equation}
1-|R|~ \propto~ k_\perp \quad \hbox{as}\quad k_\perp~\rightarrow~0\,
\end{equation}
and, consequently, $s\to 0$ and $r_{\rm inel} \to 0$.
The effect is called {\em universal 
quantum reflection}\cite{LennardJones1,LennardJones2}.

Quantum reflection can be described {\it phenomenologically\/} in an
effective single particle picture with a complex optical potential.
The {\it many-body\/} aspect of the problem is to determine the
physical origin, the magnitude, and the shape as well as possible
non--locality of that ``optical potential''.  The
energy range where quantum reflection is visible in a many-body system
like one of those considered here depends sensitively on the
energy--loss mechanisms and calls for a quantitative calculation.
Even in the limit of zero incident energy, the self energy
(\ref{selfimp}) is non-hermitian, and thus allows in principle for
sticking.  Furthermore, this energy range is
strongly affected by the long range features of the substrate
potentials\cite{BrenigSticking,BrenigSticking2,ColeQuantumReflection,%
ColeSticking}.

Specifically, in the 3-9 substrate potential models (\ref{Usubst}),
the sticking coefficient $s$ depends on the strength $C_3$
of the potential: For a local potential with the asymptotic form $C_3
z^{-3}$ as $z\to\infty$, one can show\cite{BrenigSticking2} that the
amplitude of the wave function inside the potential depends linearly
on the normal momentum of the incoming particle.  Increasing $C_3$
makes the potential appear smoother for particles with long wave
length, thus increasing the penetration depth and the probability to
reach the film.  Indeed, a calculation of the sticking coefficient
from the non-Hermitian effective Schr\"odinger equation (\ref{onedyn})
gives, already in the distorted wave Born approximation (DWBA), $s
\propto k$.

{\it Inelastic\/} scattering is, at low incident energies, only
possible by coupling to ripplons. An analysis of the imaginary part of
$\Sigma({\bf r},{\bf r}',\omega)$ reveals that the contribution of the
inelastic channels is proportional to $E^{7/2}$ which gives in the
DWBA $r_{\rm inel} \propto E^4$. In other words, inelastic processes
are negligible in the low--energy regime.

\begin{figure}[h]
 \epsfxsize=15truecm
 \centerline{\epsffile{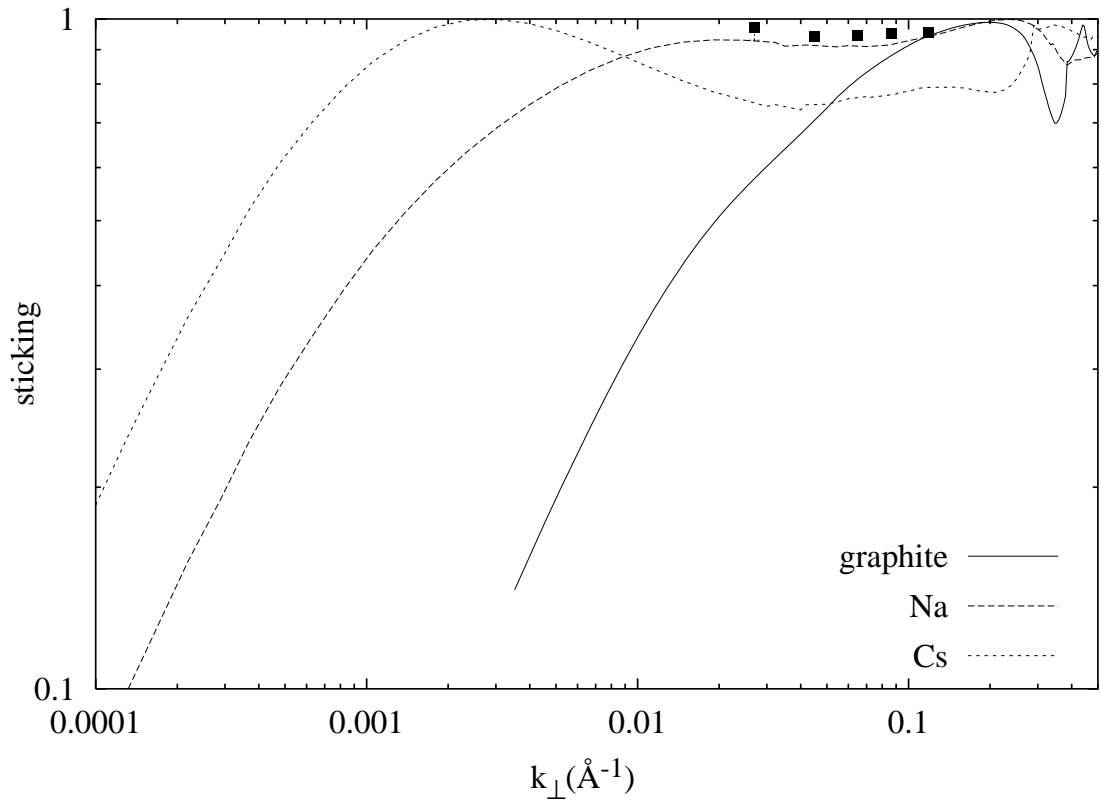}}
\vspace{2truecm}
 \caption[] {
        \label{FIGsticking1}
        Sticking on a graphite, Na, and Cs--adsorbed
        film of $n=300 A^{-2}$. The square boxes in the
        upper left of the plot are the data of Ref.
        \protect\onlinecite{Nayak83}. Note that these
        data were taken at an impact angle of 60$\deg$.
 }
\end{figure}

Although it is not the main thrust of our paper, we have examined the
low--energy reflection probabilities.  Fig. \ref{FIGsticking1} shows
three examples for the dependence of the sticking probability $s
\approx 1 - |R|^2$ on the the incident energy for normal incidence.
While on graphite
adsorbed films, quantum sticking is readily observable in the sense
that the sticking coefficient starts to drop monotonically for
wavelengths longer than 0.1$\,{\rm\AA}^{-1}$, corresponding to
energies less than 0.1~K, the linear dependence of $s$ on $k$ begins
only at energies that are two to three orders of magnitude less for Cs
adsorbed films (and similarly Mg and Na).


Once the origin and properties of the optical potential for
low--energy scattering are understood from a microscopic point of
view, one may {\it a posteriori\/} construct simple, analytic models
that provide, within the range suggested by the estimated accuracy of
the microscopic picture, some flexibility to examine the dependence of
$s$ on features of the optical potential. A simple model consists of a
local potential that approaches the substrate potential in the asymptotic
region $z\to\infty$ and that is approximated by a square well with a
depth estimated from the binding energy of the Andreev state and a
width of 15~$\rm\AA$. The energy dissipation term can be included
through a localized imaginary part of the typical magnitude of our
self--energy. Such a model reproduces qualitatively the large values
of $s$ in the mK energy regime. Of course, the model fails to explain
the dependence on ${\bf k}_\|$, see Fig. \ref{FIGRcs300log}. For
completeness, we should also add that retardation should be taken into
account for quantitative results below
1-10~mK\cite{ColeQuantumReflection}.

\subsection{Ripplon coupling}

``Quantum reflection'' as a generic phenomenon needs only {\it some\/}
damping mechanism; we now turn to the task of many--body theory to
identify and examine the physics that leads to damping. The basic
physics is contained in the self--energy (\ref{selfimp}) used in our
calculation; it includes the energy loss of an incoming particle with
energy $\hbar\omega$ to a background excitation $\hbar\omega_m$,
leaving the particle in the state $t_r$. Within this model, damping is
expected to be somewhat underestimated because the possibility to emit
two or more phonons has been neglected.

Unless there is negligible overlap of the wave functions, the most
efficient energy loss mechanism is the coupling to the lowest--lying
excitation. These lowest lying excitations of the helium film are the
surface waves ({\it ripplons}), hence one expects that the energy loss
of the $^3$He particle is dominated by the emission of a ripplon.
This serves as a {\it qualitative\/} argument. However, the reality is
more complicated for $^3$He scattering because several states are
accessible. The condition that an excitation contributes to the
imaginary part of the self--energy is that the energy denominator of
the self--energy (\ref{selfimp}) vanishes, i.e. $\hbar\omega_m + t_r =
\hbar\omega$, and there are several open channels even for vanishing
incident energy. First, the particle can, although less efficiently,
also couple to higher film excitations and can be promoted into either
the second Andreev state or into a bound state in the bulk liquid. The
reflection coefficients also depends visibly on the real part of the
self--energy, and no quantitative statement can be made without proper
treatment of both. The argument holds even at normal incidence, and
infinitesimal asymptotic energy of the impinging particle.

We show in Figs. \ref{FIGRgr300}-\ref{FIGRcs400} a few typical
examples of the reflection probability $|R({\bf k}_\|,k_\perp)|^2$ for
scattering from $^4$He films adsorbed on graphite, Na, and Cs
substrates. In contrast to experiments on atomic scattering of $^4$He
from free $^4$He surfaces\cite{Nayak83}, there is evidently a strong
dependence on the parallel wave vector ${\bf k}_\|$ which needs to be
explained in terms the possible decay channels discussed above. Since
it is unlikely that a specific feature is due to a delicate
cooperation between film and impurity degrees of freedom, it is
legitimate to discuss film-- and single particle excitations
independently.

The fact that ripplon coupling is the {\it dominant\/} energy loss
mechanism can be verified in various ways. The simplest one is the
inspection of the self--energy (\ref{selfimp}): The imaginary part of
the self--energy is, with a few exceptions to be discussed below,
localized in the surface region where the ripplon lives.  The
consequence is that, at energies below the roton, the wave functions
of the impinging particle decays basically within the surface
region. The effect can be seen in the wave functions and even
better in the probability currents which basically decay within the
surface region. A ``resonance'' in Figs. \ref{FIGgr300wav} and
\ref{FIGmg300wav} will be discussed momentarily.

From looking at Figs. \ref{FIGRgr300}-\ref{FIGRcs400}, it appears that
quantum reflection is seen only for the graphite substrate. As
explained above, this is simply a consequence of the fact that the
reflection becomes visible only at much lower energies on the alkali
metal substrates.  To demonstrate this, we have magnified in
Fig. \ref{FIGRcs300log} the low--energy region for the Cs substrate;
consistent with Fig. \ref{FIGsticking1}, it is seen that the
reflectivity starts to rise at impact energies of less than .001~K.

\subsection{Single particle resonances}

While the generic many--body aspect of all scattering and in
particular damping mechanisms must be kept in mind, one--body pictures
can occasionally --- as above for quantum reflection ---
provide useful paradigms in cases
where the process under consideration can be described in terms of the
degrees of freedom of a single particle. Such an effect is the
coupling to {\em single--particle resonances}.  A convenient and
physically illustrative definition of a resonance at an energy
$\hbar\omega$ is a large probability $|\psi_I({\bf r},\omega)|^2$ in
the region of interaction. The resulting large dissipation will render
$|R(\omega)|^2$ small.

\begin{figure}[h]
 \epsfxsize=15truecm
\vspace{2truecm}
 \centerline{\epsffile{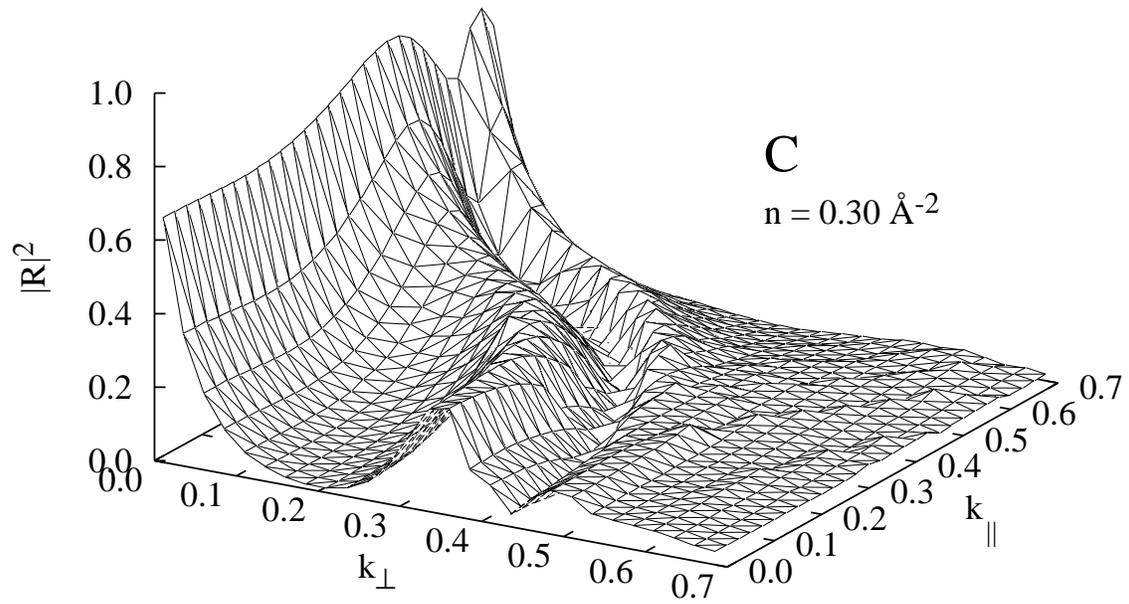}}
 \caption[Reflection coefficient on graphite-30] {
        \label{FIGRgr300}
        Dependence of reflection coefficient on wave vector magnitude
        and angle from a graphite film of density $n=0.30 A^{-2}$.
 }
\end{figure}

\begin{figure}
 \epsfxsize=15truecm
 \centerline{\epsffile{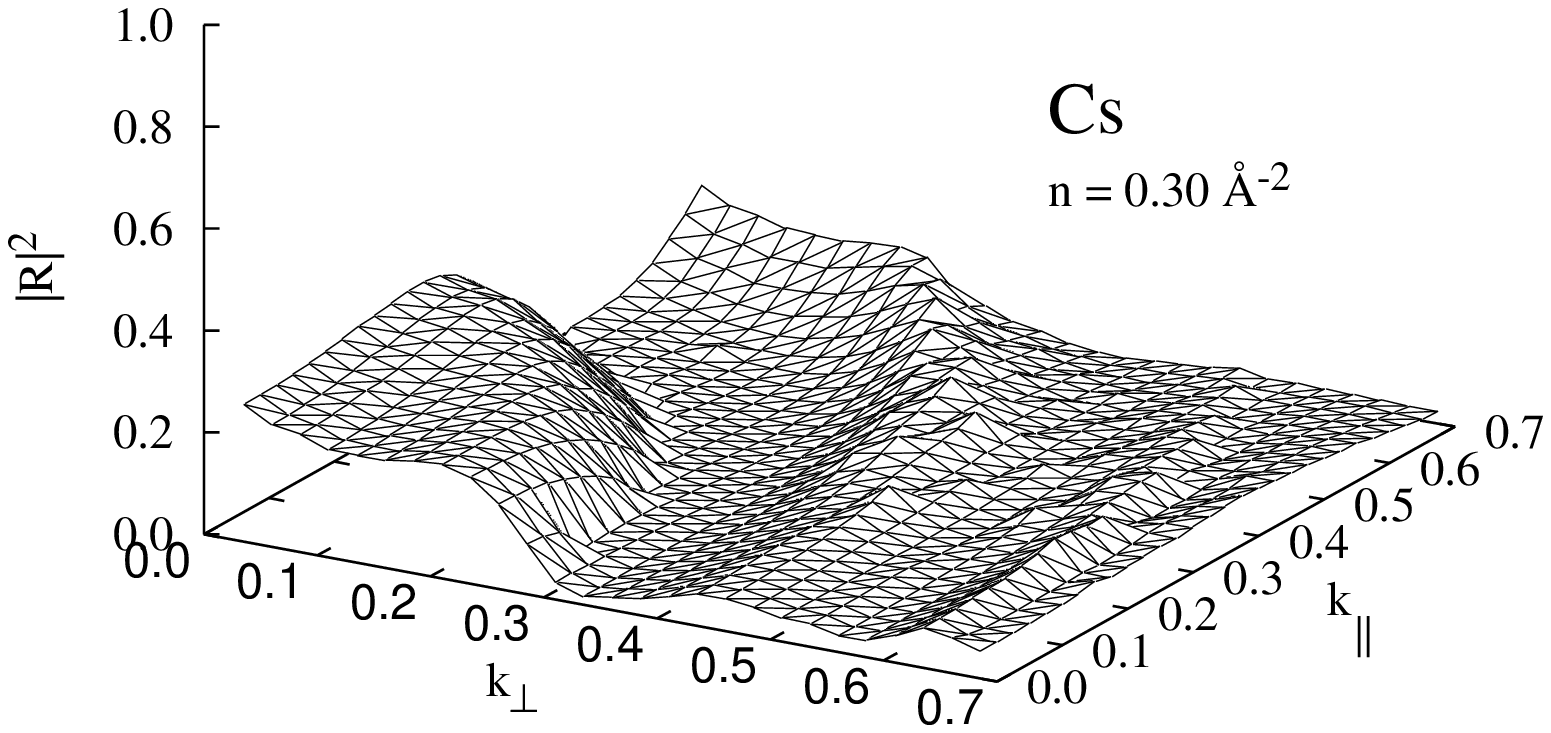}}
\vspace{2truecm}
 \caption[Reflection coefficient on Cs-30] {
        \label{FIGRcs300}
        Same as \ref{FIGRgr300} for a
        Cs film of density $n=0.30 A^{-2}$.
 }
\end{figure}

\begin{figure}
 \epsfxsize=15truecm
 \centerline{\epsffile{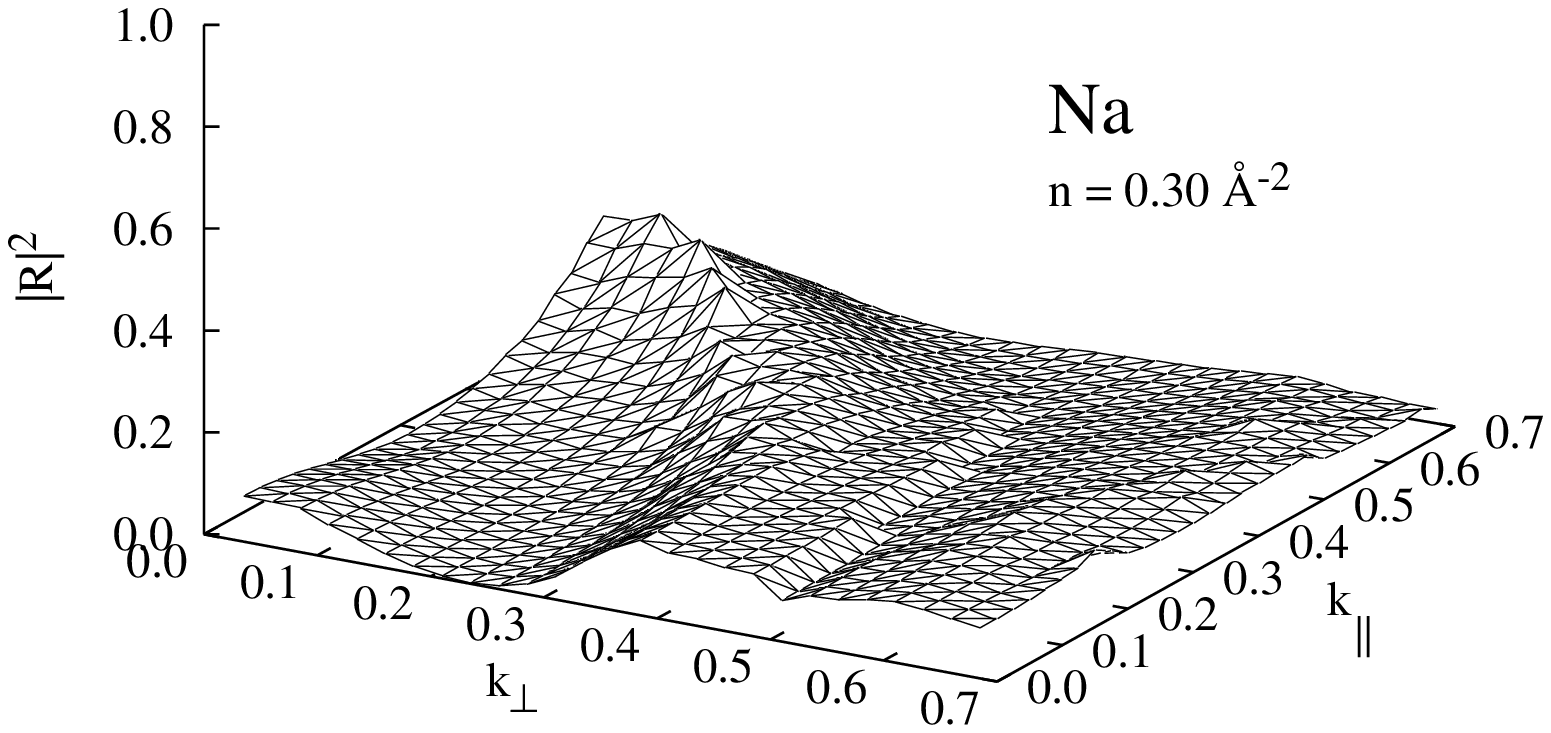}}
\vspace{2truecm}
 \caption[Reflection coefficient on Na-30]{
        \label{FIGRna300}
        Same as \ref{FIGRgr300} for a
        Na film of density $n=0.30 A^{-2}$.
 }
\end{figure}

\begin{figure}
 \epsfxsize=15truecm
 \centerline{\epsffile{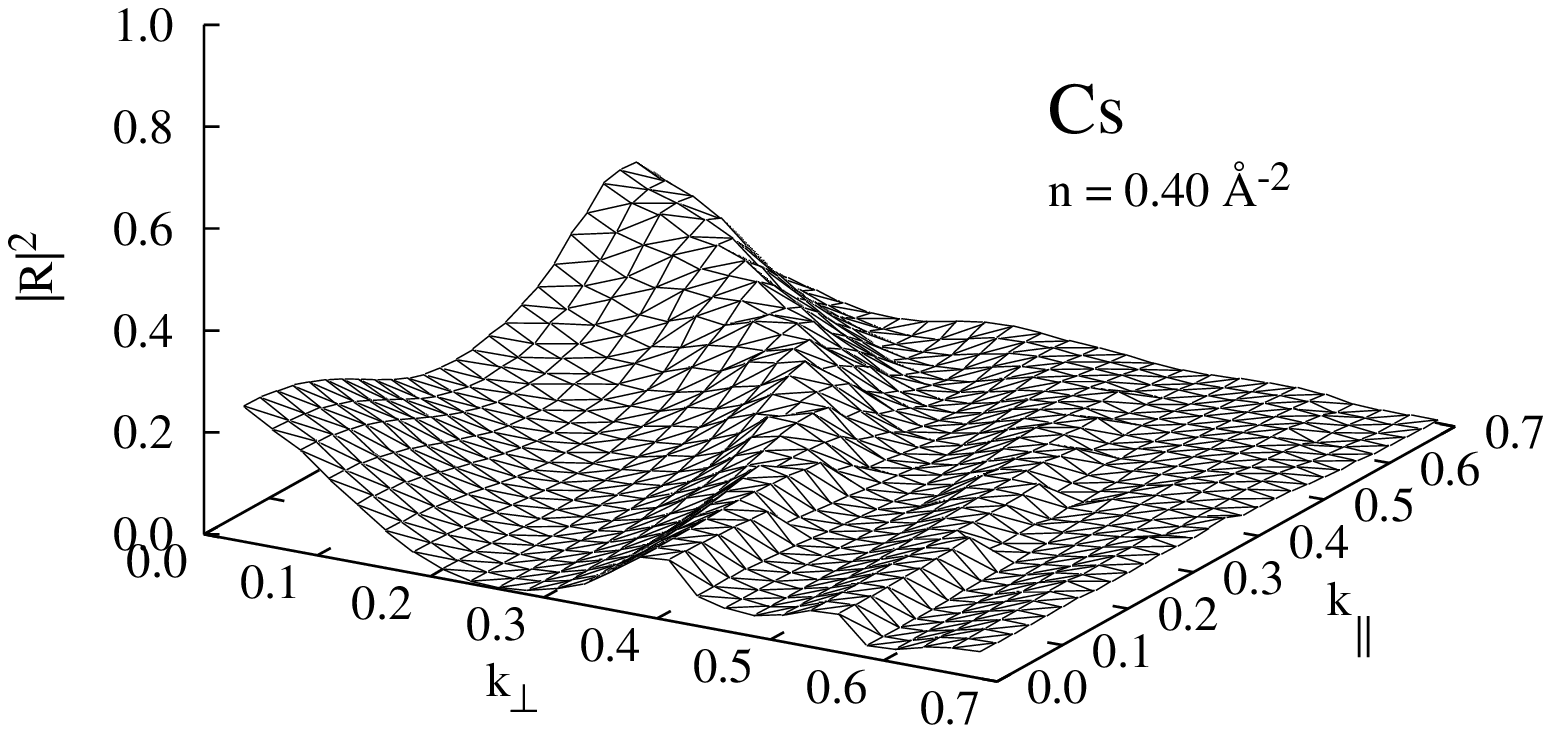}}
\vspace{2truecm}
 \caption[Reflection coefficient on Cs-40]{
        \label{FIGRcs400}
        Same as \ref{FIGRgr300} for a
        Cs film of density $n=0.40 A^{-2}$.
 }
\end{figure}

\begin{figure}
 \epsfxsize=15truecm
 \centerline{\epsffile{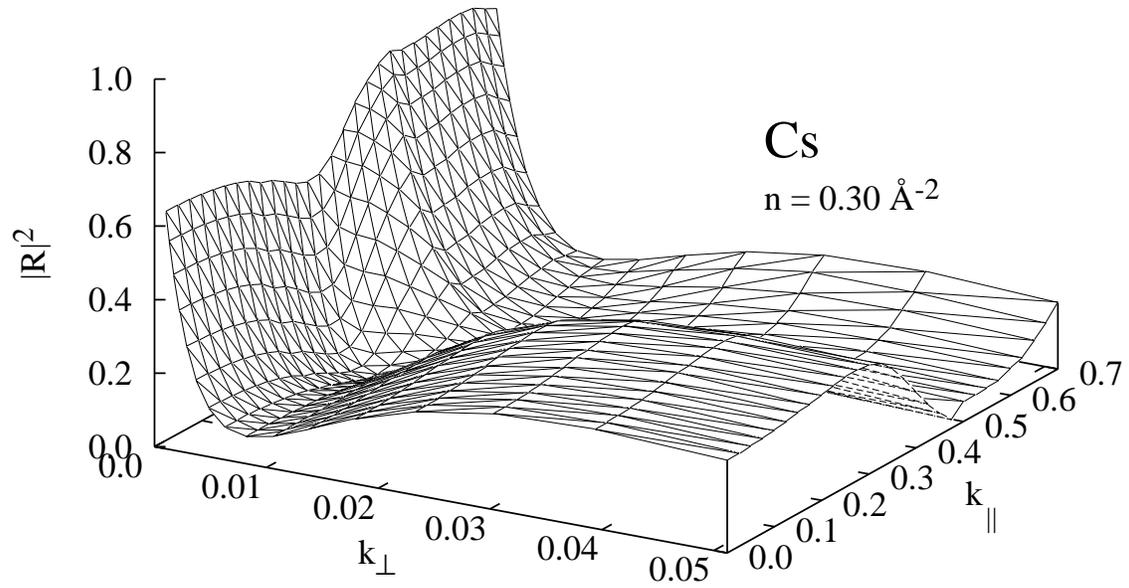}}
\vspace{2truecm}
 \caption[Reflection coefficient on Cs-30] {
        \label{FIGRcs300log}
        Fig. \ref{FIGRcs300} is magnified into the regime of low $k_\perp$
        to demonstrate that $|R|$ finally approaches unity.
 }
\end{figure}

\begin{figure}
 \epsfxsize=15truecm
\vspace{2truecm}
 \centerline{\epsffile{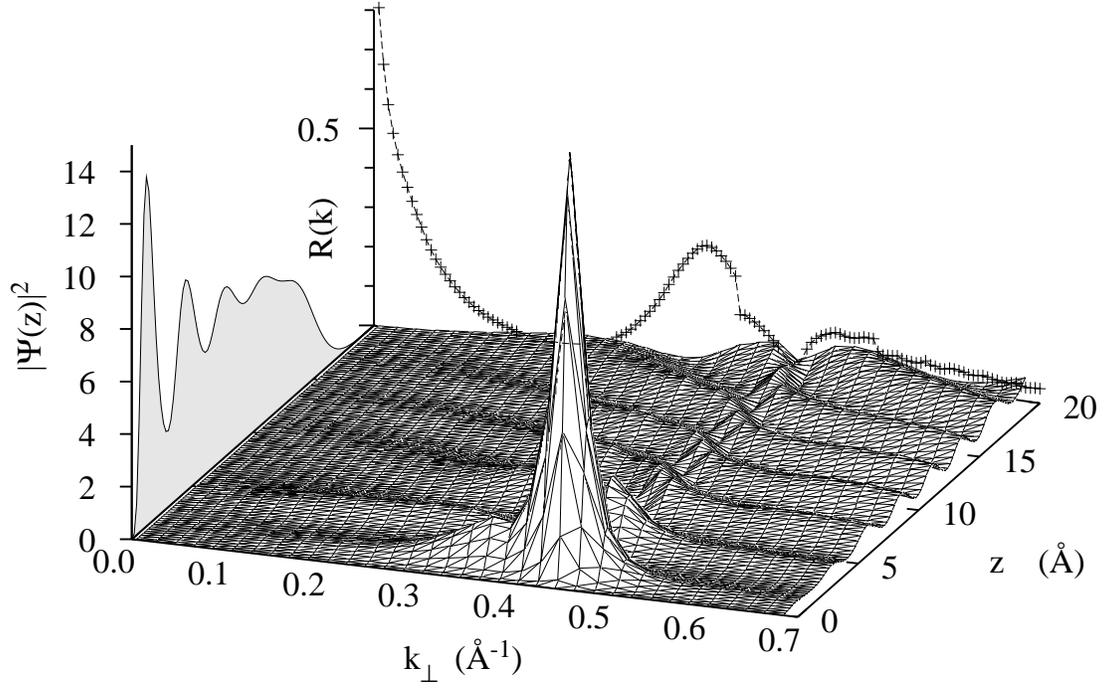}}
 \caption[substrate potentials] {
        \label{FIGgr300wav}
        The figure shows the wave function $|\psi(z)|^2$ of a
        $^3$He as a function of distance $z$ and perpendicular
        wave number $p_\perp$ for normal incidence. The left
        face shows, for reference, the density profile of the
        film and the back face the reflection coefficient
        $R(p_\perp)$. The substrate is graphite {\it plus\/}
        two solid helium layers, the surface coverage is
        $n=0.300 {\rm\AA}^{-2}$.
 }
\end{figure}

\begin{figure}
 \epsfxsize=15truecm
\vspace{2truecm}
 \centerline{\epsffile{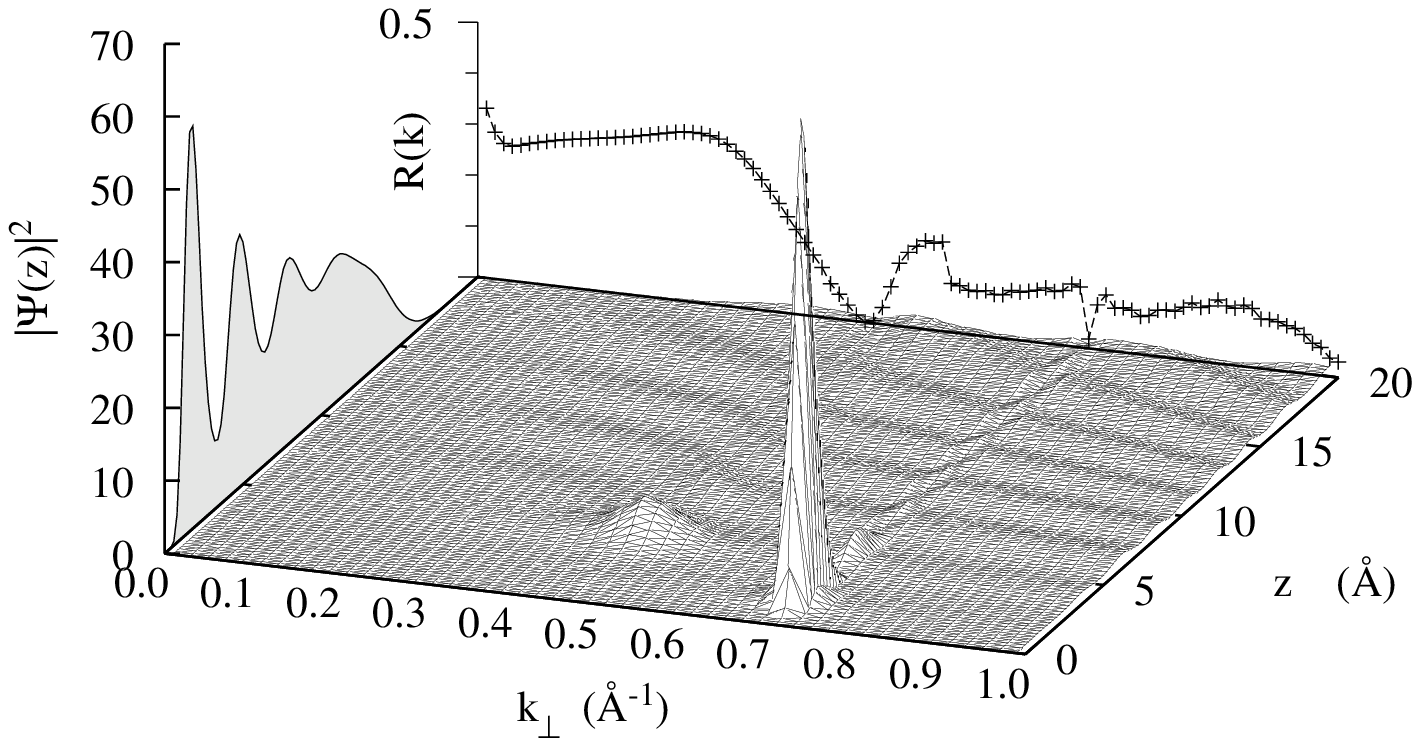}}
 \caption[substrate potentials] {
        \label{FIGmg300wav}
        Same as \ref{FIGgr300wav} for a film of $n=0.300 A^{-2}$
        on a Mg substrate.
 }
\end{figure}

The peak of the wave function close to the substrate at $k_\perp
\approx 0.4~{\rm\AA}^{-1}$ and $z \approx 1.2~{\rm\AA}$ shown in
Fig. \ref{FIGgr300wav} is a very pronounced example of such a
resonance. It displays exactly the phenomenon discussed above that the
interfacial Andreev state turns into a resonance as the substrate
strength is increased. The energy of this resonance is significantly
reduced by the coupling to {\it virtual\/} phonons: The resonance has
an energy of approximately 6~K in the static approximation
(\ref{HARimp}). Including the dynamic self--energy corrections through
the (real part of) $\Sigma(\omega)$, the resonance energy drops to
approximately 1.3~K. The energy where the wave function has a strong
peak in the vicinity of the substrate coincides with that of the dip in
the reflection coefficient. Fig.  \ref{FIGRgr300} shows this for the
special case of zero parallel momentum, but the agreement between the
peak of the wave function and the minimum of the reflection
probability persists at all parallel momenta. Also seen clearly in
Fig. \ref{FIGgr300wav} is the change in the phase of the wave as the
resonance is crossed as a function of energy.

The elliptic ridge of the reflection coefficient as a function of
($k_\perp$, $k_\|$) can be explained by the coupling of the
interfacial Andreev resonance discussed above to the virtual
excitations of the film. This has the consequence that the resonance
acquires an effective mass\cite{filmdirt} $m^*_{\rm res}$.  At zero
parallel momentum, the position of the dip in the reflection
coefficient agrees with the location of the resonance seen in
Fig. \ref{FIGgr300wav}. The shape of the ridge can be explained by
assuming that {\it all\/} of the energy of the impinging particle is
deposited in that resonance. Energy conservation and momentum
conservation {\it parallel\/} to the substrate then leads to the
relationship
\begin{equation}
\epsilon_{\rm res} = {\hbar^2 k_\perp^2\over 2m_3}
+ {\hbar^2 k_\|^2\over 2m_3}\left[1-{m_3\over m^*_{\rm res}}\right]\,,
\label{resmass}
\end{equation}
where $\epsilon_{\rm res}$ is the energy of the resonance. Following
the peak of the wave function in the resonance in the $(k_\perp,
k_\|)$ plane leads, within the accuracy that can be expected from such
a relatively crude argument, to the same conclusion. Basically -- and
expectedly --- the same resonances occur at other surface coverages;
the precise location of the dip in the reflection varies due to the
multitude of other open scattering channels. A similar resonance
occurs in the more strongly attractive Mg substrates, the
corresponding wave functions are shown in Fig. \ref{FIGmg300wav}. In
this case, one finds a second resonance in the second layer which is,
however, less pronounced. A list of energies and effective masses is
given, together with the values for the Andreev state and the results
of Ref. \onlinecite{filmdirt} of the bound states, in Table
\ref{Table1}. The effective masses were obtained by fitting the curve
defined by Eq. (\ref{resmass}) to reproduce the location of the peak
of the wave function within the visible region. As pointed out above,
the weaker substrate Cs has a bound state to which the scattering
particle cannot couple, whereas the Na substrate is a marginal case.

In all cases considered here we have found a significant dependence of
the reflection coefficient $\left|R\right|$ on the {\it parallel\/}
momentum, {\it cf.\/} Figs. \ref{FIGRgr300}-\ref{FIGRcs400}. Such a
feature can not be explained within a local, complex single--particle
model, is not also not seen in experiments on $^4$He scattering
off $^4$He films/surfaces\cite{Nayak83}. The feature is most
pronounced on graphite and Mg substrates, {\it cf.\/}
Fig. \ref{FIGRgr300}.

\begin{figure}
 \epsfxsize=15truecm
\vspace{2truecm}
 \centerline{\epsffile{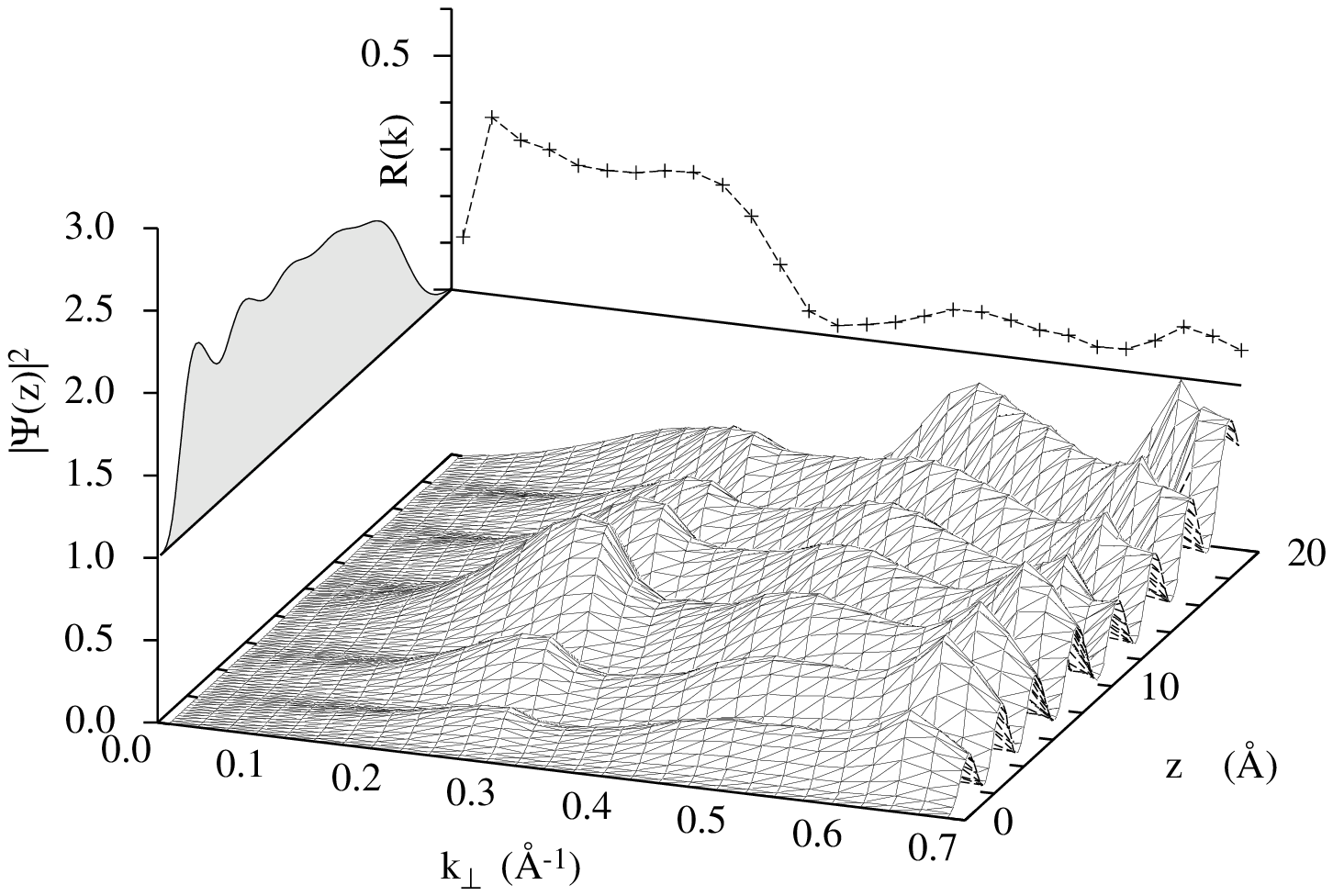}}
 \caption[substrate potentials] {
        \label{FIGcs300wav}
        Same as \ref{FIGgr300wav} for a film of $n=0.300 A^{-2}$
        on a Cs substrate.
 }
\end{figure}

\begin{figure}
 \epsfxsize=15truecm
 \centerline{\epsffile{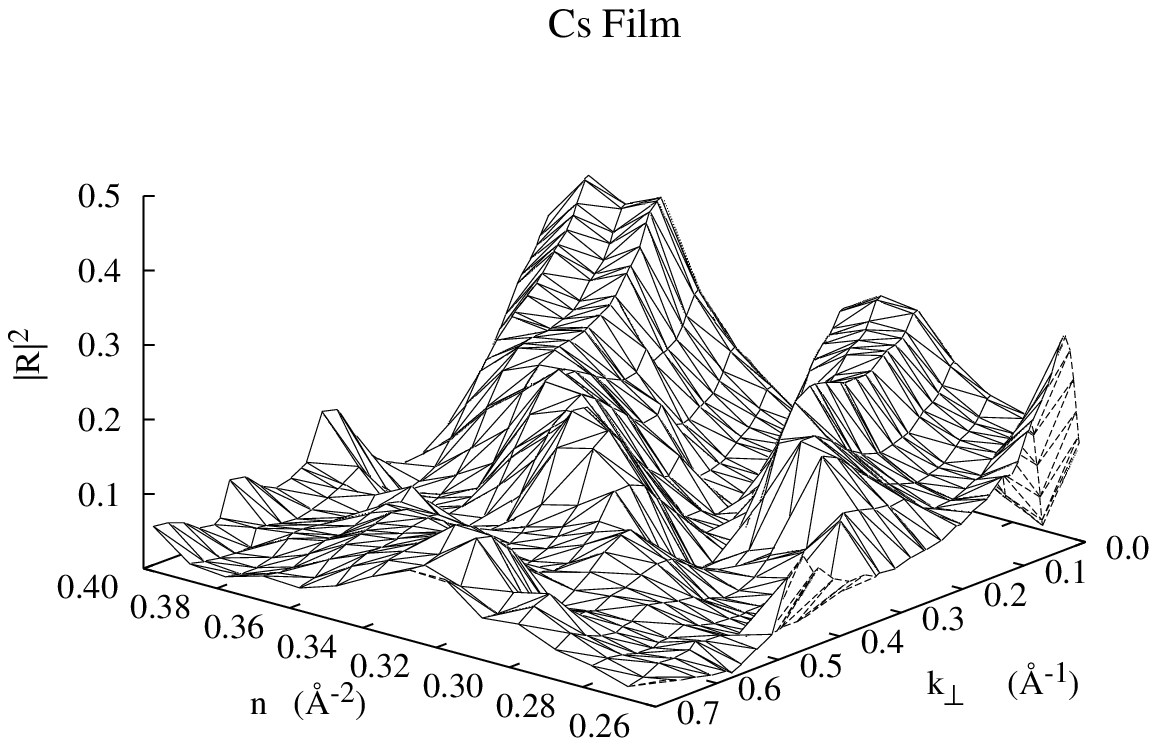}}
\vspace{2truecm}
 \caption[roton coupling] {
        \label{fig:csRofn}
        Reflection coefficient $|R|^2$ for normal incidence on
        a a sequence films with surface coverages between
        $n=0.26{\rm \AA}^{-2}$ and $n=0.39{\rm \AA}^{-2}$
        on a Cs-adsorbed film. 
 }
\end{figure}

Also for the other substrates (see Figs.
\ref{FIGRcs300}, \ref{FIGRna300} and \ref{FIGRcs400}), $R(\omega)$
depends on the parallel component of the momentum.  One sees similar,
but broader ridges in the reflection coefficient, but {\it no\/} sharp
peaks in the wave function, {\it cf.\/} Fig. \ref{FIGcs300wav}. The
effect can also be explained by the features of the impurity states
within the film. But this time, the impurity states are {\it not\/}
localized but are extended states that will, with increasing film
thickness, develop to $^3$He states dissolved in the $^4$He
liquid. Consistent with this picture, the energy of the resonances in
Na and Cs adsorbed films decreases with increasing surface coverage
$n$, until they become bound states, {\it cf.} \ref{fig:csRofn}.
This slipping below the threshold
$E=0$ is best seen in the phase shift at $E\to 0$, which jumps
whenever this occurs (in case of Cs: at $n=0.330$ and $0.380{\rm
\AA}^{-2}$).

\subsection{Roton coupling}

The presence of roton excitations affects the scattering properties
at two levels. First, at {\it all\/} energies, the coupling to
virtual rotons is a significant contribution to the {\it real part\/}
of the self--energy; omitting these contributions by, for example,
restricting the state sums in the self--energy (\ref{selfimp})
to energies below the roton minimum, leads to reflection coefficients
that are, even at energies well below the roton minimum, about a
factor of 2 smaller than when excitations in the roton regime
are included. This is to some extent plausible since the roton
is a reflection of the short--range structure of the system which
is dominated by the core repulsion, and such effects should make
the film look ``stiffer''.

At higher energies, the coupling to roton excitations also opens a new
damping mechanism. As seen in Fig. \ref{fig:skwcs40o}, ``roton--like''
excitations appear also for film excitations perpendicular to the
surface, and the impinging particle can couple to these
excitations. In our calculations, the effect of roton coupling should
become visible at an energy of about 15~K, this is because we have
used a Feynman-spectrum in the energy denominator of
Eq. (\ref{selfimp}). In previous work\cite{filmexc}, we have scaled
the energy denominators in the self-energy by an amount such that the
roton is placed at roughly the right energy. We have refrained from
this phenomenological modification since this procedure
would also scale the ripplon away from its correct value which
is already obtained in the Feynman approximation.

Above 15~K, the film loses its elastic reflectivity for $^3$He atoms
completely. There is, of course, still the possibility of
some inelastic scattering, but we consider this scenario
unlikely from our experience with the propagation of
$^3$He impurities in {\it bulk\/} $^4$He\cite{SKJLTP}. Hence, we
expect that $^3$He atoms will be completely absorbed by $^4$He films
when the impact energy is above the energy of the bulk roton. The
effect is also seen quite clearly in the wave function of the
scattering particle which does not penetrate into the film at all at
energies above that of the roton.

\section{Summary}

We have set in this paper the basic scenario for calculations of atom
scattering processes from inhomogeneous $^4$He. This work parallels
similar research on scattering of $^4$He atoms\cite{scatlett} and also
provides the groundwork for applications on the presently active
area of atom scattering from $^4$He clusters.

Technically, the calculations presented here are somewhat simpler than
those for $^4$He atom scattering\cite{scatlett} (since the Hartree impurity
spectra appearing in the energy denominators in \ref{selfimp} are
decoupled in parallel and perpendicular motion in contrast to the
ripplon/phonon/roton spectra) which enabled us to do a systematic
study of the dependence of the reflection coefficient on the
parallel momentum.

When applicable, our general conclusions are very similar to those
that we have drawn for $^4$He atom scattering: Most of the physics
happens due to ripplon coupling, the wave function is substantially
damped in the surface region. ``Quantum reflection''
does not come to bear until energies as low as $0.1~K$ on a graphite
substrate, and $.01~K$ or less on alkali metals. A second
damping mechanism happening at higher energies is the coupling to
rotons, this effect dampens the impurity motion completely; an
equivalent effect is expected, and found, in bulk $^4$He.

A new aspect specific to $^3$He scattering is the coupling to
single--particle resonances within the film; such an effect will not
be seen for $^4$He scattering. We have demonstrated that the
properties of the remaining reflected particles are directly
influenced by the features of the impurity states within these films
and that scattering experiments can directly measure the energy and
the ``effective mass'' of these resonances. Fully acknowledging the
experimental difficulty of the task, we hope that these findings will
inspire further measurements on $^3$He scattering off $^4$He surfaces
and films.

Unfortunately it is difficult to make direct comparisons with
experiments available today\cite{Nayak83}. One reason is that we are
with our calculations apparently still too far from the bulk limit
that a comparison is meaningful. This is most clearly seen in the
oscillatory dependence of the reflection coefficient on the energy of
the incoming particle even in a case if a relatively thick film
without localized resonances within the film
(Fig. \ref{FIGcs300wav}). There is also still pronounced non-monotonic
dependence of the reflection coefficient on the surface coverage, {\it
cf.\/} Fig. \ref{fig:csRofn}.

Further applications of our
work are twofold: One is the application to scattering of hydrogen
isotopes off $^4$He surface. While experimental efforts in this area
have been significantly stronger\cite{Berkhout86,Doyle91,YuDoyle93},
the situation is less rich: The H impurity is only very weakly bound
and can lose its energy only to the ripplon, in other words the
imaginary part of the self--energy (\ref{selfimp}) comes from one
state only. Moreover, the H atom must overcome a potential barrier of
about 10~K to penetrate into the bulk liquid which makes the coupling
to any interior degrees of freedom negligible.

Similarly interesting is the possibility of scattering experiments of
both $^3$He and $^4$He atoms off $^4$He {\it droplets.\/} These
experiments, to be carried out in the energy regime of a few tenth to
a few degrees, would also couple to both surface and volume modes and
should also more clearly display the coupling to excitations inside
the droplets. In the spherical geometry, inelastic processes of the
kind described here cannot occur at low energy because
the continuous quantum number ${\bf k}_\|$ is replaced by the discrete
angular momentum. Hence, all low-lying modes are discrete and the
energy denominator of Eq.  (\ref{selfimp}) will normally be non--zero,
in other words the self-energy is hermitean. Second, that these
systems are not contaminated by substrate effects and should therefore
allow for a cleaner interpretation of the results. We have learned
that such scattering experiments have meanwhile been
performed\cite{ToenniesPrivate} and show indeed the expected coupling
of $^3$He particles to roton-like excitations inside the droplets.

Such experiments and calculations would also provide an ideal scenario
to test ideas and procedures established in nuclear physics in a much
better controlled and --- in terms of the underlying Hamiltonian ---
better understood physics.

Calculations in this direction are in progress and will be published
elsewhere.

\section{ACKNOWLEDGEMENTS}

This work was supported, in part, by the Austrian Science Fund under
project P11098-PHY. We thank S. E. Campbell, R. B. Hallock, J. Klier,
and M. Saarela for valuable discussions.

\newpage

\bibliography {papers}

\begin{thebibliography}{10}

\bibitem{WyattReflectivity2}
A.~F.~G. Wyatt, M.~A.~H. Tucker, and R.~F. Cregan, Phys. Rev. Lett. {\bf 74},
  5236  (1995).

\bibitem{WyattReflectivity1}
M.~A.~H. Tucker and A.~F.~G. Wyatt, J. Low Temp. Phys. {\bf 100},  105  (1995).

\bibitem{WyattSpectrum}
M.~A.~H. Tucker and A.~F.~G. Wyatt, Physica B {\bf 194-196},  549  (1994).

\bibitem{WyattRough}
R.~F. Cregan, M.~A.~H. Tucker, and A.~F.~G. Wyatt, J. Low Temp. Phys. {\bf
  101},  531  (1995).

\bibitem{Edwards75}
D.~O. Edwards {\it et~al.}, Phys. Rev. Lett. {\bf 34},  1153  (1975).

\bibitem{EdwardsFatourosScattTh}
D.~O. Edwards and P.~P. Fatouros, Phys. Rev. B {\bf 17},  2147  (1978).

\bibitem{Nayak83}
V.~U. Nayak, D.~O. Edwards, and N. Masuhara, Phys. Rev. Lett. {\bf 50},  990
  (1983).

\bibitem{SwansonEdwardsScatt4He}
D.~R. Swanson and D.~O. Edwards, Phys. Rev. B {\bf 37},  1539  (1988).

\bibitem{EdwardsFatourosHe3Scatt}
D.~O. Edwards, P.~P. Fatouros, and G.~G. Ihas, Physics Letters A {\bf 59},  131
   (1976).

\bibitem{Godfried85}
H.~P. Godfried {\it et~al.}, Phys. Rev. Lett {\bf 55},  1311  (1985), measuring
  of adsorbtion energies and recombination rate.

\bibitem{Berkhout86}
J.~J. Berkhout, E.~J. Wolters, R. van Roijen, and J.~T.~M. Walraven, Phys. Rev.
  Lett. {\bf 57},  2387  (1986).

\bibitem{Doyle91}
J.~M. Doyle, J.~C. Sandberg, I.~A. Yu, and et~al, Phys. Rev. Lett. {\bf 67},
  603  (1991).

\bibitem{YuDoyle93}
I.~A. Yu, J.~M. Doyle, J.~C. Sandberg, and et~al, Phys. Rev. Lett. {\bf 71},
  1589  (1993).

\bibitem{ColeSticking}
C. Carraro and M.~W. Cole, Phys. Rev. B {\bf 45},  12930  (1992).

\bibitem{ColeQuantumReflection}
C. Carraro and M.~W. Cole, Z. Phys. B {\bf 98},  319  (1995), diese Ausgabe von
  Z. Phys. enthaelt die Proceedings on Ions and Atoms in Superfluid Helium
  1994.

\bibitem{Bittner94}
E.~R. Bittner and J.~C. Light, J. Chem. Phys. {\bf 102},  2614  (1995).

\bibitem{filmstruc}
B.~E. Clements, J.~L. Epstein, E. Krotscheck, and M. Saarela, Phys. Rev. B {\bf
  48},  7450  (1993).

\bibitem{filmdyn}
B.~E. Clements {\it et~al.}, Phys. Rev. B {\bf 50},  6958  (1994).

\bibitem{filmexc}
B.~E. Clements, E. Krotscheck, and C.~J. Tymczak, Phys. Rev. B {\bf 53},  12253
   (1996).

\bibitem{filmdirt}
B.~E. Clements, E. Krotscheck, and M. Saarela, Phys. Rev. B {\bf 55},  5959
  (1997).

\bibitem{Impu}
E. Krotscheck, M. Saarela, and J.~L. Epstein, Phys. Rev. B {\bf 38},  111
  (1988).

\bibitem{KSE}
E. Krotscheck, M. Saarela, and J.~L. Epstein, Phys. Rev. Lett. {\bf 61},  1728
  (1988).

\bibitem{EKS}
J.~L. Epstein, E. Krotscheck, and M. Saarela, Phys. Rev. Lett. {\bf 64},  427
  (1990).

\bibitem{scatlett}
C.~E. Campbell, E. Krotscheck, and M. Saarela, Phys. Rev. Lett  (1997),
  submitted.

\bibitem{HalleyDrops}
J.~W. Halley, C.~E. Campbell, C.~F. Giese, and K. Goetz, Phys. Rev. Lett. {\bf
  71},  2429  (1993).

\bibitem{ChuckWoodsLT20}
C.~E. Campbell and J.~W. Halley, Physica B {\bf 194-196},  533  (1994).

\bibitem{SettyHalleyCampbell97}
A. Setty, J.~W. Halley, and C.~E. Campbell, Phys. Rev. Lett. {\bf 79},  3930
  (1997).

\bibitem{KohnSticking}
D.~P. Clougherty and W. Kohn, Phys. Rev. B {\bf 46},  4921  (1993).

\bibitem{BrenigSticking}
W. Brenig, Z. Physik B {\bf 36},  227  (1980).

\bibitem{BrenigSticking2}
J. B{\"o}lheim, W. Brenig, and J. Stuzki, Z. Physik B {\bf 48},  43  (1982).

\bibitem{Brivio94}
G.~P. Brivio, T.~B. Grimley, and G. Guerra, Surf. Sci. {\bf 320},  344  (1994).

\bibitem{OldOldSurf}
E. Krotscheck, Phys. Rev. B {\bf 32},  5713  (1985).

\bibitem{FeenbergBook}
E. Feenberg, {\em Theory of Quantum Liquids} (Academic, New York, 1969).

\bibitem{ChangCohen}
C.~C. Chang and M. Cohen, Phys. Rev. A {\bf 8},  1930  (1973).

\bibitem{ZarembaKohn77}
E. Zaremba and W. Kohn, Phys. Rev. B {\bf 15},  1769  (1977).

\bibitem{Cole}
M.~W. Cole, D.~R. Frankl, and D.~L. Goodstein, Rev. Mod. Phys. {\bf 53},  199
  (1981).

\bibitem{Nacher}
P.~J. Nacher and J. Dupont-Roc, Phys. Rev. Lett. {\bf 67},  2966  (1991).

\bibitem{Ketola}
K.~S. Ketola, S. Wang, and R.~B. Hallock, Phys. Rev. Lett. {\bf 68},  201
  (1992).

\bibitem{Taborek}
J.~E. Rutlege and P. Taborek, Phys. Rev. Lett. {\bf 68},  2184  (1992).

\bibitem{KLW96}
J. Klier and A.~F.~G. Wyatt, Czekoslowak Journal of Physics Suppl. {\bf 46},
  439  (1996).

\bibitem{Taborek95}
D. Ross, P. Taborek, and J.~E. Rutlege, Phys. Rev. Lett. {\bf 74},  4483
  (1995).

\bibitem{alkalis}
B.~E. Clements, E. Krotscheck, and C.~J. Tymczak, J. Low Temp. Phys. {\bf 107},
   387  (1997).

\bibitem{LauterExeter}
H.~J. Lauter, H. Godfrin, V.~L.~P. Frank, and P. Leiderer,  in {\em Excitations
  in Two-Dimensional and Three-Dimensional Quantum Fluids}, Vol.~257 of {\em
  NATO Advanced Study Institute, Series B: Physics}, edited by A.~F.~G. Wyatt
  and H.~J. Lauter (Plenum, New York, 1991), pp.\ 419--427.

\bibitem{LauterPRL}
H.~J. Lauter, H. Godfrin, V.~L.~P. Frank, and P. Leiderer, Phys. Rev. Lett.
  {\bf 68},  2484  (1992).

\bibitem{LauterJLTP}
H.~J. Lauter, H. Godfrin, and P. Leiderer, J. Low Temp. Phys. {\bf 87},  425
  (1992).

\bibitem{DevilleLT21}
G. Deville, P. Roche, N.~J. Appleyard, and F.~I.~B. Williams, Czekoslowak
  Journal of Physics Suppl. {\bf 46},  89  (1996).

\bibitem{Deville97}
P. Roche, G. Deville, N.~J. Appleyard, and F.~I.~B. Williams, J. Low Temp.
  Phys. (rapid communications) {\bf 106},  565  (1997).

\bibitem{Surface2}
E. Krotscheck, Phys. Rev. B {\bf 31},  4258  (1985).

\bibitem{Feynman}
R.~P. Feynman, Phys. Rev. {\bf 94},  262  (1954).

\bibitem{SKJLTP}
M. Saarela and E. Krotscheck, J. Low Temp. Phys. {\bf 90},  415  (1993).

\bibitem{Hallock94}
D.~T. Sprague, N. Alikacem, P.~A. Sheldon, and R.~B. Hallock, Phys. Rev. Lett.
  {\bf 72},  384  (1994).

\bibitem{Hallock89}
R.~H. Higley, D.~T. Sprague, and R.~B. Hallock, Phys. Rev. Lett. {\bf 63},
  2570  (1989).

\bibitem{Hallock88}
J.~M. {Valles, Jr.}, R.~H. Higley, R.~B. Johnson, and R.~B. Hallock, Phys. Rev.
  Lett. {\bf 60},  428  (1988).

\bibitem{EdSaam}
D.~O. Edwards and W.~F. Saam,  in {\em Progress in Low Temperature Physics},
  edited by D.~F. Brewer (North Holland, New York, 1978), Vol.~7A, pp.\
  282--369.

\bibitem{LennardJones1}
J.~E. Lennard-Jones, F.~R. Strachan, and A.~F. Devonshire, Proc. R. Soc. London
  {\bf 156},  6  (1936).

\bibitem{LennardJones2}
J.~E. Lennard-Jones, F.~R. Strachan, and A.~F. Devonshire, Proc. R. Soc. London
  {\bf 156},  29  (1936).

\bibitem{ToenniesPrivate}
J. Harms and P. Toennies, 1998, (private communication).

\end{thebibliography}
\bibliographystyle{prsty}
\newpage
\begin{table}
\setdec 0.00
\caption{Resonance energies and effective masses of the
interfacial Andreev state on various substrates.  The first line
gives, for reference, the data of the Andreev state at the free
surface, and the second the interfacial Andreev state on a Cs
substrate (From Ref. \protect\onlinecite{filmdirt}).  The last three
lines give the results obtained here from scattering properties.
Energies are given in K, the row labeled with ``C'' refers to the
graphite {\it plus\/} two solid layers of $^4$He model used in this
work.}
\bigskip
\begin{tabular}{lcc}
Substrate &Energy & $m^*/m_3$\\
\tableline
-  & \dec -5.4 &  1.3\\
Cs & \dec -4.3 & 1.7\\
C  & \dec 1.3$\pm$0.3 & 1.7$\pm$0.3\\
Mg & \dec 4.3$\pm$0.5 & 1.6$\pm$0.2 \\
\end{tabular}
\label{Table1}
\end{table}
\newpage
%

%
%

\newpage
\newpage


\newpage


\newpage


\newpage


\newpage

\newpage
\newpage
%
%
\newpage
%
%
\newpage
%
%
\newpage
%
%
%
%
\newpage
\newpage
%
%
\newpage
%
%
\newpage
%
%
\newpage
%
%
\begin{figure}[h]
 \epsfxsize=15truecm
 \centerline{\epsffile{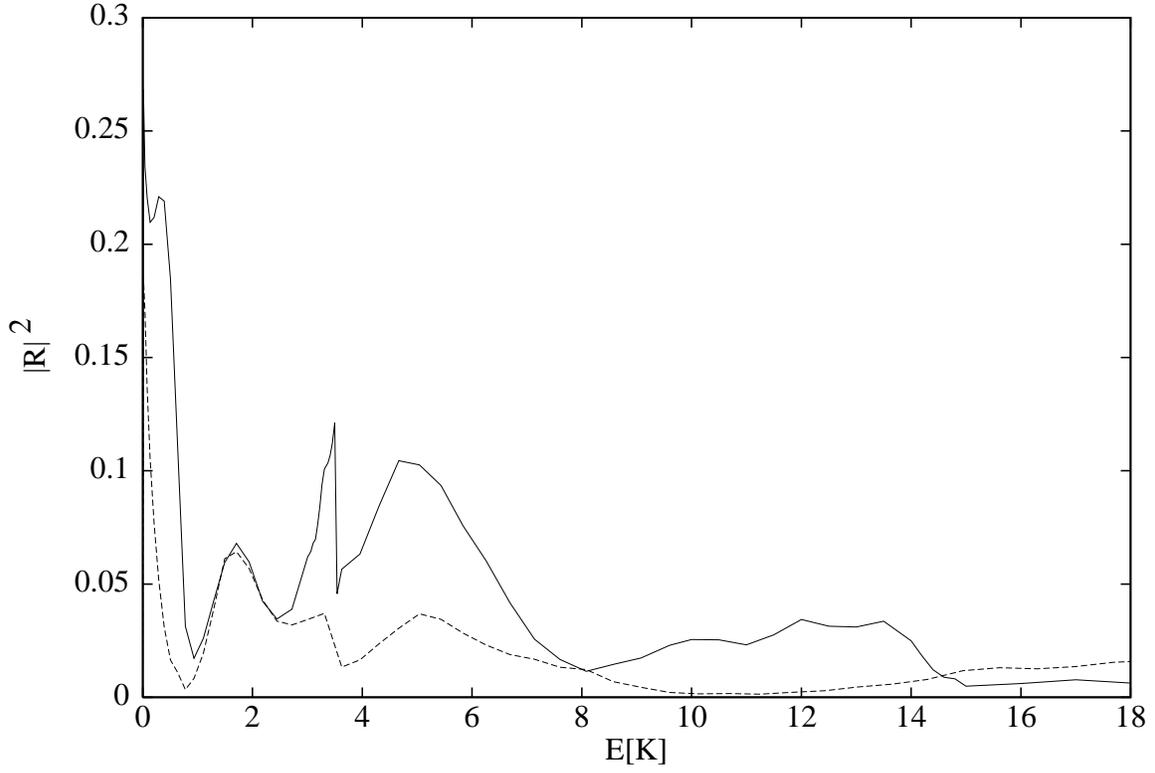}}
\vspace{2truecm}
 \caption[roton coupling] {
        \label{FIGcs300rotons}
        Reflection coefficient $|R|^2$ for normal incidence on
        a film with $n=0.300{\rm \AA}^{-2}$ on a Cs-adsorbed film. 
        The solid line is the result when all relevant intermediate
        states are kept in the state sums (\protect\ref{selfimp})
        whereas the dashed line is the result when the sum over
        intermediate states is truncated below the roton minimum.
        A new scattering channel opens at 3.5~K.
 }
\end{figure}
\newpage
%
%

\end{document}